\documentclass[a4paper,11pt]{article}
\usepackage{jheppub} 
\usepackage{bbm}
\usepackage{tikz-feynman}
\usepackage{slashed}
\usepackage{cleveref}
\crefformat{equation}{Eq.~(#2#1#3)}
\crefmultiformat{equation}{Eqs.~(#2#1#3)}{ and~(#2#1#3)}{, (#2#1#3)}{ and~(#2#1#3)}
\crefformat{figure}{Fig.~(#2#1#3)}
\crefformat{section}{Sec.~#2#1#3}
\crefformat{table}{Tab.~#2#1#3}
\usepackage{float}
\usepackage{cancel}
\usetikzlibrary{cd}
\usetikzlibrary{positioning}
\tikzset{tail reversed/.code={\pgfsetarrowsstart{tikzcd to}}}
\usepackage{adjustbox}

\newcommand{\PP}{\frac{1 + \slashed{v}}{2}}

\newcommand{\MS}{\overline{\text{MS}}}

\newcommand{\fdfrac}[2]{\mbox{\footnotesize$\displaystyle\frac{#1}{#2}$}}

\makeatletter
\gdef\@fpheader{\vspace{2ex}}
\makeatother

\title{\boldmath Position-space renormalization schemes for four-quark operators in HQET}

\author[1,2]{Joshua Lin,}
\author[1,2]{William Detmold,}
\author[3]{Stefan Meinel}

\affiliation[1]{Center for Theoretical Physics, Massachusetts Institute of Technology, Cambridge, MA 02139, USA}
\affiliation[2]{The NSF AI Institute for Artificial Intelligence and Fundamental Interactions}
\affiliation[3]{Department of Physics, University of Arizona, Tucson, AZ 85721, USA}

\emailAdd{joshlin@mit.edu}
\emailAdd{wdetmold@mit.edu}
\emailAdd{smeinel@arizona.edu}

\preprint{ \vbox{\hbox{MIT-CTP/5707 }}}

\abstract{
$X$-space schemes are gauge-invariant, regulator-independent renormalization schemes that are defined by requiring position-space correlation functions of gauge-invariant operators to be equal to their noninteracting values at particular kinematic points. 
These schemes can be used to nonperturbatively renormalize composite operators in Lattice Quantum Chromodynamics (LQCD), and by computing matching coefficients between the $X$-space scheme and $\MS$ in the dimensionally-regulated continuum, matrix elements calculated with LQCD can be converted to $\MS$-renormalized matrix elements. 
Using $X$-space schemes for Heavy Quark Effective Theory (HQET) operators has the additional benefit that appropriate ratios of position-space correlation functions cancel the power-divergent static-quark self-energy of Lattice HQET nonperturbatively. 
This work presents the $O(\alpha_S)$ matching coefficients between $X$-space renormalized four-quark flavor-nonsinglet HQET operators relevant for the lifetimes of charm- and bottom-hadrons, and four-quark HQET operators relevant for mixing between neutral mesons containing a heavy quark, such as $B-\overline{B}$ mixing.  
}

\begin{document}
\maketitle
\flushbottom

\section{Introduction}
\label{sec:intro}

Lattice Quantum Chromodynamics (LQCD) calculations are crucial to nonperturbative determinations of Standard-Model parameters and searches for beyond-the-Standard-Model physics in the flavor sector \cite{USQCD:2019hyg}. 
These calculations (see FLAG \cite{FlavourLatticeAveragingGroupFLAG:2021npn} for a review) typically require renormalization of composite operators, as the hadronic matrix-elements are renormalization scale-dependent, and must be combined with Wilson coefficients to determine renormalization-scale-independent physical quantities. 
The most commonly used renormalization scheme in phenomenological applications of QCD is the modified minimal-subtraction ($\MS$) scheme because of its perturbative simplicity. However, LQCD has no direct access to the $\MS$ scheme which is only defined in dimensionally-regulated perturbation theory. 
Lattice perturbation theory \cite{Capitani:2002mp} can be used to convert bare matrix-elements computed in LQCD to $\MS$-renormalized matrix-elements, but tends to suffer from poor convergence properties \cite{Gockeler:2010yr}. 
Nonperturbative renormalization schemes bypass these problems by renormalizing composite lattice operators in an intermediate, regulator-independent scheme, before perturbatively matching between this intermediate scheme and the $\MS$ scheme using dimensional regularization. 
A common choice of an intermediate scheme is to impose a momentum-space renormalization condition, as is done in the Regularization-Independent Momentum Subtraction (RI-(S)MOM) methods \cite{Martinelli:1994ty,Sturm:2009kb}. A drawback of RI-(S)MOM schemes is that gauge-fixing is required due to the use of gauge-noninvariant quark and/or gluon states, giving rise to additional mixing with gauge-noninvariant operators \cite{Bhattacharya:2015rsa, Lytle:2018evc}. Furthermore, Gribov copies (the discrete set of intersections between a gauge orbit and a gauge-fixing condition) introduce a systematic error to numerically computed renormalization constants, although numerical studies often suggest that this is practically negligible compared to statistical noise \cite{Giusti:2002rn,Paciello:1992gy}. 

To circumvent these problems, one can instead impose renormalization conditions based on position-space correlation functions of gauge-invariant operators. In the literature this has been called the $X$-space scheme \cite{Gimenez:2004me}, or the gauge-invariant-renormalization-scheme (GIRS) \cite{Costa:2021iyv}. 
For multiplicatively renormalizable operators, the $X$-space renormalized operator $\mathcal{O}^{(X)}$ can be related to the bare operator $O^{(0)}$ by $O^{(X)} = Z_\mathcal{O}^{(X)} \mathcal{O}^{(0)}$, where $Z_\mathcal{O}^{(X)}$ is an (often divergent) renormalization constant. 
One possible  $X$-space renormalization scheme can be defined by the requirement that the renormalized two-point correlation function built from $\mathcal{O}^{(X)}$ is equal to its non-interacting (NI) value when the operators are separated by a fixed spacetime distance $x$:
\begin{equation}\label{eq:1}
\big\langle \mathcal{O}^{\dagger (X)}\left(0\right) \mathcal{O}^{(X)}\left(x\right) \big\rangle = \big\langle \mathcal{O}^{\dagger(0)}\left(0\right) \mathcal{O}^{(0)}\left(x\right) \big\rangle \big|_\mathrm{NI} \ ,
\end{equation}
where $\sqrt{x^2}$ is the scale at which the operator is renormalized, and the non-interacting correlation function is defined as the $\alpha_S \to 0$ value of the correlation function\footnote{The non-interacting correlation function in $X$-space schemes is sometimes referred to as `tree-level', though this is a misnomer due to the fact that the noninteracting calculation of position-space correlation functions of composite operators involves loops. Note that the superscript labels `$(0)$' on the operators in the RHS of \cref{eq:1} are redundant as the non-interacting value for the bare and renormalised operators is the same.}. Although the scheme is gauge invariant by definition, it requires more-complicated perturbative calculations in order to match to other schemes in the continuum. Matching calculations from $X$-space schemes to $\MS$ have been performed for light-quark bilinears \cite{Chetyrkin:2010dx}, heavy-light quark bilinears \cite{Chetyrkin:2021qvd}, dimension-5 operators appearing in the energy-momentum tensor \cite{Costa:2021iyv}, and supercurrent operators in supersymmetric theories~\cite{Bergner:2022see}. 

When implementing RI-(S)MOM renormalization conditions at momentum $p$ with a lattice discretisation, there is a `window problem' where $\Lambda_\mathrm{QCD} \ll p \ll a^{-1}$ is required to keep all systematic uncertainties under control. 
Indeed, the RI-SMOM scheme \cite{Sturm:2009kb} was introduced to remove infrared convergence issues in applications of the original RI-MOM scheme where some momenta were not in the desired range. 
Here, $a$ is the lattice discretisation-scale that regulates the ultraviolet (UV) behaviour of the theory and $\Lambda_\mathrm{QCD}$ is the typical QCD scale that emerges through dimensional transmutation. 
The same window problem affects any position-space scheme, where $\sqrt{x^2} \ll \Lambda_\mathrm{QCD}^{-1}$ is required to control perturbation-theory errors appearing in the perturbative matching to $\MS$, and $a \ll \sqrt{x^2}$ is required to control discretisation artifacts. In practice, this window problem must be investigated on a case-by-case basis, and various investigations have been performed in $X$-space schemes for the local light-quark bilinear operators using Wilson fermions \cite{Gimenez:2004me,Cali:2020hrj}, twisted-mass fermions \cite{Cichy:2012is}, as well as domain-wall fermions \cite{Tomii:2016xiv}. 
Furthermore, numerical studies of the feasibility of $X$-space renormalization conditions for renormalizing the QCD energy-momentum tensor \cite{Spanoudes:2022gow}, heavy-light quark bilinear operators \cite{Korcyl:2015xmd}, and operators in supersymmetric field theories \cite{Soler:2022bfj} have been undertaken. There have also been numerical investigations of the possibility of using position-space schemes to match between three and four-flavor QCD \cite{Tomii:2019esd,Tomii:2020smd}. 

The $X$-space scheme is particularly suited to renormalizing Heavy Quark Effective Theory (HQET) operators. Choosing a reference frame in which the heavy-quark velocity has spatial components that vanish, $v = (1,0,0,0)^T$, the bare Euclidean propagator for a heavy quark $Q$ in the static limit is naturally written in position space \cite{Eichten:1989zv} as
\begin{equation}\label{eq:hqprop}
\langle Q^{(0)}(0) \overline{Q}^{(0)}(x_E) \rangle_F = \delta_{\vec{x}_E,\vec{0}} \ \theta(-t_E)  W^{(0)}(0,x_E) \frac{1 + \gamma_0}{2},
\end{equation}
where $\langle \cdot \rangle_F$ indicates the path integral is performed over all the fermionic degrees of freedom but not the gauge degrees of freedom, and $W^{(0)}(a,b)$ is the bare straight Wilson line from $a$ to $b$. A complication in lattice regularizations of the static theory is that the static-quark self-energy has a power divergence \cite{Sommer:2015hea}, which is caused by mixing between the kinetic term $\overline{Q} D_0 Q$ and a mass-like term $m_\mathrm{stat} \overline{Q} Q$, where $m_\mathrm{stat} \sim O(\alpha_S)/a$ is radiatively generated.  
The $X$-space scheme proposed in this work (\Cref{sec:db0,sec:db2}) utilizes ratios of three-point to two-point position-space correlation functions to nonperturbatively cancel this power divergence. Nonperturbatively-renormalized matrix-elements of HQET operators can thus be extracted without needing to determine $m_\mathrm{stat}$ explicitly, which would otherwise constitute another source of uncertainty. 

In this work, a set of $X$-space schemes for renormalizing four-quark HQET operators is proposed, and the $O(\alpha_S)$ matching to $\MS$ is calculated in the static limit, extending the $X$-space approach used in Refs.~\cite{Gimenez:2004me,Costa:2021iyv,Bergner:2022see}. 
The first set of operators that are considered are isospin-nonsinglet, four-quark operators $\tau_{f f'} (\overline{Q} \Gamma q_f)(\overline{q}_{f'} \Gamma' Q)$ where $\Gamma,\Gamma'$ are spin-colour tensors, $q_f \in \{q_u,q_d\}$ are light-quark fields, and $\tau_{f f'}$ is a Pauli matrix in the light-quark flavor space. In the Heavy Quark Expansion (HQE) formalism, matrix elements of these $\Delta Q = 0$ (heavy-quark flavor preserving) operators are known as `spectator contributions', and are the dominant $O(1/m_Q^3)$-corrections to the inclusive lifetimes of heavy (charm or bottom) hadrons \cite{Neubert:1996we,Lenz:2014jha}. 
The second set of operators that are considered are $\Delta B = 2$ four-quark operators, relevant for determinations of $B -\overline{B}$-mixing \cite{Gabbiani:1996hi}. 
Precise determinations of hadronic matrix-elements of these quantities will allow for better constraints on fundamental parameters of the Standard Model such as the CKM matrix-elements, and will also further constrain extensions of the Standard Model. 
Existing LQCD studies of these four-quark operators have utilized lattice perturbation-theory to perform matching to $\MS$ \cite{Ishikawa:2011dd,DiPierro:1998ty,DiPierro:1999tb,Gimenez:1998mw}, and the nonperturbative renormalization-conditions proposed in this work will allow for more precise LQCD determinations of the renormalized matrix-elements \cite{Lin:2022fun}. 

The four-quark operators studied in this work do not renormalize multiplicatively but rather mix within multiplets of operators with the same quantum numbers, and therefore \cref{eq:1} must be generalized to define a scheme that determines the entire mixing-matrix. 
In order to provide a sufficient number of renormalization conditions to determine the full mixing-matrix, the $X$-space scheme proposed here utilize three-point correlation functions involving the four-quark operator with multiple choices of source and sink operators. In particular, the source and sink operators used are heavy-light mesonic operators $\overline{Q} \Gamma q$ and heavy-light-light baryonic operators $\epsilon^{abc} [q^{aT} \Gamma_1 q^b] \Gamma_2 Q^c$ for various choices of the Dirac matrices $\Gamma,\Gamma_1,\Gamma_2$. 
Perturbative calculations of two-point correlation functions constructed from these operators have been performed in the literature, and these can be used to determine matching coefficients between $X$-space schemes and the $\MS$ scheme, which are presented to $O(\alpha_S)$ in \Cref{sec:source}. 
The four-quark operators also mix into evanescent operators
(operators that explicitly vanish in $d = 4$) in dimensional regularization. 
To be able to utilize the $X$-space scheme as a regulator-independent scheme for conversion of lattice matrix-elements, the scheme is defined in terms of evanescent-subtracted operators. 
The $X$-space schemes, and $O(\alpha_S)$-matching to the $\MS$ scheme for the isospin non-singlet $\Delta B = 0$ operators and the $\Delta B = 2$ operators are presented in \Cref{sec:db0,sec:db2}, respectively. An outlook is presented in \Cref{sec:conc}.
An overview of the conventions and integration techniques used in this paper is given in \Cref{sec:app}.

\section{Multiplicatively Renormalizable Operators}
\subsection{$X$-space schemes for heavy-light bilinear and heavy-light-light trilinear operators}
\label{sec:source}

In this section, the $O(\alpha_S)$ matching-factors between $X$-space-renormalized operators and $\MS$-renormalized operators for heavy-light mesonic, and heavy-light-light baryonic operators in HQET are presented. Choosing the frame in which the static heavy quark propagates purely in the Euclidean time direction, the Euclidean HQET Lagrangian used in the following calculations is given by
\begin{equation}\label{eq:lag}
\mathcal{L} = \frac{1}{4} F_{\mu \nu} F_{\mu \nu} + \sum_{f = \{u,d\}} \overline{q}_f \gamma_\mu D_\mu q_f +  \overline{Q} D_0 Q,
\end{equation}
with two massless light-quarks $q_u$ and $q_d$, and one static heavy-quark $Q$ that satisfies $\frac{1 + \gamma_0}{2} Q = Q$. 
To regulate the continuum theory, dimensional regularization (DR) is used, where the dimension of spacetime is analytically continued to $d = 4 - \epsilon$, and $\gamma_5 =  \gamma_0 \gamma_1 \gamma_2 \gamma_3$ is treated in the 't Hooft-Veltman (HV) scheme \cite{tHooft:1972tcz} (see \Cref{app:charge} for a review of the conventions used). 
Due to the heavy-quark term in the action not containing any Dirac matrices, in $d = 4$ there is an $SU(2)_h$ heavy-quark spin symmetry
\begin{equation}
Q \mapsto e^{-i\theta_j \gamma_5  \gamma_j} Q, \quad \overline{Q} \mapsto \overline{Q} e^{i\theta_j \gamma_5  \gamma_j} 
\end{equation}
that leaves the action invariant, where $\gamma_j \in \{\gamma_1,\gamma_2,\gamma_3\}$. 
The resulting effect of this symmetry is that local heavy-light mesonic operators are related by heavy-quark symmetry in the following $SU(2)_h$ doublets:
\begin{equation}\label{eq:mesdoub}
SU(2)_h \begin{cases}
\ H_{f}^+ (0^+) : \overline{q}_f Q \\
{H}_{f,i}^{*+} (1^+) : \overline{q}_f \gamma_5 \gamma_i Q
\end{cases},\ \ 
SU(2)_h \begin{cases}
\ H_f^- (0^-) : \overline{q}_f  \gamma_5 Q \\
{H}_{f,i}^{*-} (1^-) : \overline{q}_f  \gamma_i Q
\end{cases},
\end{equation}
where the $J^P$ quantum numbers of the state are listed along with the corresponding local heavy-light operator. Note that the $\pm$ superscripts refer to the parity of the operator, rather than the electromagnetic charge of the state. In \Cref{eq:mesdoub} and what follows, $\gamma_i$ indexes the $(d-1)$-dimensional spatial Dirac matrices, such that $\gamma_i \gamma_i = (d-1) \mathbbm{1}$.\footnote{In $d = 4$, where spatial Dirac matrices $\gamma_i$ take values in $\gamma_i \in \{\gamma_1,\gamma_2,\gamma_3\}$, the heavy-quark spin symmetry is indeed $SU(2)_h$. In dimensional regularization the spatial index $i$ now varies over the $(d-1)$-dimensions that are not timelike, and formally the symmetry group is no longer $SU(2)_h$. This distinction, however, does not have any implications for the calculations presented in this section. } 
Since the antiparticle is integrated out in HQET, the operators shown in \cref{eq:mesdoub} form a basis for the heavy-light bilinear operators with no derivatives in $d = 4$. 

The heavy-light-light baryonic operators of the form $\epsilon^{abc} [q^{aT} \tau \Gamma_1 q^b] \Gamma_2 Q^c$ for varying Dirac matrices $\Gamma_1,\Gamma_2$ and isospin matrices $\tau$ are also multiplicatively renormalizable. In the following, $C$ is the charge-conjugation matrix satisfying $C \gamma_\mu C^{-1} = -\gamma_\mu^T$, with an explicit construction in dimensional regularization given in \Cref{app:charge}. The isospin matrices are written in terms of the antisymmetric $\tau^A$ or symmetric $\tau^S_\alpha$ matrices
\begin{equation}
\tau^A := \begin{pmatrix} 0 & 1 \\ -1 & 0 \end{pmatrix}, \quad \tau^S_1 := \begin{pmatrix} 1 & 0 \\ 0 & 0 \end{pmatrix}, \quad  \tau^S_0 := \frac{1}{\sqrt{2}}\begin{pmatrix} 0 & 1 \\ 1 & 0 \end{pmatrix}, \quad \tau^S_{-1} := \begin{pmatrix} 0 & 0 \\ 0 & 1 \end{pmatrix}.
\end{equation}
Operators coupling to $\Lambda_Q$ baryons (isospin singlets) are given by

\hspace*{-2.9cm}\vbox{\begin{align*}
\Lambda_1 \left(\tfrac{1}{2}^+\right)&: \epsilon^{abc} [q^{aT} \tau^A C  \gamma_5 q^b] Q^c , \\
\Lambda_2 \left(\tfrac{1}{2}^+ \right)&: \epsilon^{abc} [q^{aT} \tau^A C  \gamma_5 \gamma_0 q^b] Q^c , \\ 
\Lambda_1^- \left( \tfrac{1}{2}^-\right) &: \epsilon^{abc} [q^{aT} \tau^A C q^b] Q^c ,
\end{align*}}\vspace{-0.6cm}
\begin{equation}\label{eq:lam}
SU(2)_h \begin{cases}
\ \Lambda_2^- \left(\tfrac{1}{2}^-\right): \epsilon^{abc} [q^{aT} \tau^A C \gamma_5 \gamma_i q^b] \gamma_i \gamma_5 Q^c ,\\
\Lambda^{*-}_i \left( \tfrac{3}{2}^- \right) : \epsilon^{abc} [q^{aT} \tau^A C \gamma_5 \gamma_i q^b] Q^c  - \tfrac{1}{3} \epsilon^{abc} [q^{aT} \tau^A C \gamma_5 \gamma_j q^b] \gamma_i \gamma_j Q^c .
\end{cases}
%\end{split}
\end{equation}
The operators are labelled by their angular momentum representation and parity, $J^P$. 
The $SU(2)_h$-doublet $\{\Lambda_2^-, \Lambda_i^{*-}\}$ arises from decomposing the tensor product of a spin-$1$ light-quark doublet with the spin-$\frac{1}{2}$ heavy quark into irreducible spin representations. In particular, spin-$\frac{3}{2}$ operators such as $\Lambda_i^{*-}$ satisfy the condition $\gamma_i \Lambda_i^{*-} = 0$ in $d = 4$. 
Operators coupling to $\Sigma_Q$-baryons which transforms in the isospin-triplet representation, are given by

\vspace{0.3cm}
\hspace*{-1.45cm}\vbox{\begin{equation*}
SU(2)_h \begin{cases}
\hspace{0.18cm} \Sigma_{1,\alpha} \left(\tfrac{1}{2}^+\right): \epsilon^{abc} [q^{aT} \tau^S_\alpha C \gamma_i q^b] \gamma_i \gamma_5 Q^c ,\\
\Sigma_{1,\alpha,i}^* \left( \tfrac{3}{2}^+ \right) : \epsilon^{abc}[q^{aT} \tau^S_\alpha C \gamma_i q^b] Q^c  - \tfrac{1}{3}  \epsilon^{abc} [q^{aT} \tau_A C \gamma_j q^b] \gamma_i \gamma_j Q^c ,
\end{cases}
\end{equation*}}

\vspace{-0.4cm}\hspace*{-1.1cm}\vbox{\begin{equation*}
\hspace{0.02cm}SU(2)_h \begin{cases}
\hspace{0.21cm} \Sigma_{2,\alpha} \left(\tfrac{1}{2}^+\right): \epsilon^{abc} [q^{aT} \tau^S_\alpha C \gamma_0 \gamma_i q^b] \gamma_i \gamma_5 Q^c ,\\
\hspace{0.025cm} \Sigma_{2,\alpha,i}^* \left( \tfrac{3}{2}^+ \right) : \epsilon^{abc} [q^{aT} \tau^S_\alpha C \gamma_0 \gamma_i q^b] Q^c  - \tfrac{1}{3}  \epsilon^{abc} [q^{aT} \tau^S_\alpha C \gamma_0 \gamma_j q^b] \gamma_i \gamma_j Q^c ,
\end{cases}
\end{equation*}}

\vspace{-1cm}\begin{equation*}
\hspace{0.02cm}SU(2)_h \begin{cases}
\hspace{0.18cm} \Sigma_{1,\alpha}^- \left(\tfrac{1}{2}^-\right): \epsilon^{abc} [q^{aT} \tau^S_\alpha C \gamma_5 \gamma_0 \gamma_i q^b] \gamma_i \gamma_5 Q^c, \\
\Sigma_{1,\alpha,i}^{*-} \left( \tfrac{3}{2}^- \right) : \epsilon^{abc}[q^{aT} \tau^S_\alpha C \gamma_5 \gamma_0 \gamma_i q^b] Q^c  - \tfrac{1}{3}  \epsilon^{abc}[q^{aT} \tau^S_\alpha C \gamma_5 \gamma_0 \gamma_j q^b] \gamma_i \gamma_j Q^c ,
\end{cases}
\end{equation*}
\begin{equation}\label{eq:sig}
\hspace{-4.8cm}
\begin{aligned}\hspace{0.15cm}\Sigma_{2,\alpha}^- \left(\tfrac{1}{2}^-\right): \epsilon^{abc} [q^{aT} \tau^S_\alpha C  \gamma_0 q^b] Q^c , 
\end{aligned}
\end{equation}
where $\alpha$ is the isospin index and $i$ is the vector index of the spin-$\frac{3}{2}$ field. The operators in \cref{eq:lam,eq:sig} form a complete basis in $d = 4$ for local HQET heavy-light-light currents with no derivatives, of which the positive-parity local operators were previously classified in Refs~\cite{Groote:1996xb,Grozin:1992td}. 

The $X$-space scheme for multiplicatively-renormalizable currents is specified at a renormalization scale $t^{-1}$ by the following condition:
\begin{equation}\label{eq:Xcond}
\left\langle \mathcal{O}^{\dagger (R,X)}\left(-t,\vec{0}\right) \mathcal{O}^{(R,X)}\left(t,\vec{0}\right) \right\rangle = \left\langle \mathcal{O}^{\dagger(0)} \left(-t,\vec{0}\right) \mathcal{O}^{(0)}\left(t,\vec{0}\right) \right\rangle \bigg|_\mathrm{NI},
\end{equation}
where $R$ is a regulator (either $\mathrm{DR}$ or the lattice regulator), and $X$ specifies the operators are renormalized in the $X$-space scheme. The subscript NI on the right-hand side of \Cref{eq:Xcond} refers to the non-interacting value of the correlator. If the operator has additional indices (such as isospin triplet indices $\alpha$ in the case of the $\Sigma$-interpolating operators in \cref{eq:sig}, or spatial gamma-matrix indices $i$) these additional indices are summed over on both sides of the renormalization condition. 
For example, in the $1^-$ mesonic channel (the $B^*$ mesons if $Q$ is a bottom quark), the renormalization condition is
\begin{equation}\label{eq:ang}
\sum_i \left\langle (\overline{q} \gamma_i Q)^{\dagger (R,X)}\left(-t,\vec{0}\right) (\overline{q} \gamma_i Q)^{(R,X)}\left(t,\vec{0}\right) \right\rangle = \sum_i \left\langle (\overline{q} \gamma_i Q)^{(0)\dagger} \left(-t,\vec{0}\right) (\overline{q} \gamma_i Q)^{(0)} \left(t,\vec{0}\right) \right\rangle \bigg|_\mathrm{NI}.
\end{equation}
Furthermore, the open Dirac indices of the static heavy-quark and heavy anti-quark in the baryonic operators are traced over. 

\begin{figure}[t]
\begin{center}
\includegraphics{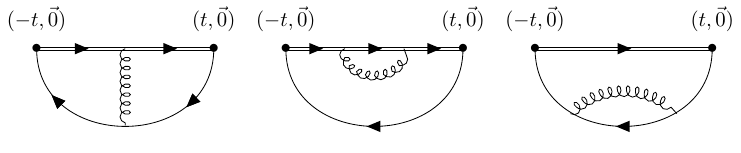}
\end{center}
  \caption{Diagrams for the $O(\alpha_S)$ contributions to the two-point position space correlator shown in \Cref{eq:Xcond}, when $\mathcal{O}$ is one of the heavy-light currents listed in \Cref{eq:mesdoub}. The operator insertions are represented by the filled in dots in the diagrams above. Feynman rules are derived from \Cref{eq:lag}, double lines refer to the static-quark propagator, single lines are light-quark propagators, and curly lines are gluons. The unlabelled positions of the interaction vertices are integrated over. }\label{fig:BB}
\end{figure}

The position-space diagrams that contribute to the $O(\alpha_S)$ determination of the two-point correlation function for the heavy-light currents are shown in Fig.~\ref{fig:BB}. 
Although chiral symmetry is broken by the HV-prescription for $\gamma_5$, the massless nature of the light quark still has consequences for the symmetries present in the renormalization factors. 
In particular, the specific $\Gamma$-matrix appearing in the interpolating operator $\mathcal{O} = \overline{q} \Gamma Q$ only affects the two-point correlation function by a constant factor, as the correlation function (to any order in $\alpha_S$) has the form
\begin{align}\label{eq:BBstruc}
\left\langle \left( \overline{Q}^{(\mathrm{DR},0)} \overline{\Gamma} q^{(\mathrm{DR},0)}\right)(-t,\vec{0}) \ \left(\overline{q}^{(\mathrm{DR},0)} \Gamma Q^{(\mathrm{DR},0)}\right)(t,\vec{0}) \right\rangle &= A \ \mathrm{Tr}\left(\frac{1 + \gamma_0}{2} \overline{\Gamma} \gamma_0 \Gamma \right) \nonumber \\
&= A P \ \mathrm{Tr}\left(\frac{1 + \gamma_0}{2} \overline{\Gamma} \Gamma \right),
\end{align}
where $A$ is a constant that does not depend on $\Gamma$, $P \in \{-1,+1\}$ is the parity of the state ($\gamma_0\Gamma = P \Gamma \gamma_0)$, and $\overline{\Gamma} = \gamma_0 \Gamma^\dagger \gamma_0$. This is because of the fact that regardless of the number of gluons attached to the light-quark line in the two-point correlation function, there are always an odd number of $\gamma$-matrices inserted (one for each massless propagator and one for each vertex). For example, shown below is the case with three gluons attached to the light quark:
\begin{center}
\includegraphics{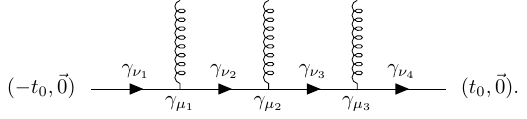}
\end{center}
 The only four-vector available for contraction is the purely timelike heavy-quark velocity $v$; thus, after performing integration over loop momenta and Fourier transforming to position space, the light-quark line is proportional to $\slashed{v} = \gamma_0$. 
The result of \Cref{eq:BBstruc} is that all heavy-light operators $\overline{q} \Gamma Q$ renormalize multiplicatively with the same factor to all orders of $\alpha_S$ in the $X$-space scheme, regardless of the choice of $\Gamma$. 

The diagrams in Fig.~\ref{fig:BB} can be computed by first calculating the corresponding two-loop momentum-space diagrams (also known as $p$-type integrals \cite{Grozin:2003ak}) and taking a Fourier transform; further details of this computation are given in \Cref{subsec:int}. Writing $(\overline{q} \Gamma Q)^{(\mathrm{DR},S)} = Z_{(\overline{q} \Gamma Q)}^{(\mathrm{DR},S)} (\overline{q}^{(0)} \Gamma Q^{(0)})$ for the heavy-light currents (where $S$ is a choice of renormalization scheme), the $X$-space renormalization factors are given by\footnote{$O(\alpha_S^2)$ calculations of the two-point position-space correlation function of heavy-light currents can be found in Ref~\cite{Chetyrkin:2021qvd}.}
\begin{equation}\label{eq:BBDRX}
Z_{(\overline{q} \Gamma Q)}^{(\mathrm{DR},X)} (t,\mu) =  1 - \frac{\alpha_S(\mu)}{ \pi \epsilon} - \frac{\alpha_S(\mu)}{\pi} \left( \frac{4}{3} + \frac{2 \pi^2}{9} + \frac{1}{2}\log\left(4 \pi e^{\gamma_E} \mu^2 t^2\right) \right)
\end{equation}
in $d = 4 - \epsilon$ dimensions. Here, $\mu$ is the scale obtained after writing the renormalized coupling-constant as a dimensionless object $g^{(\mathrm{DR},\MS)} = \frac{1}{Z_g}{\mu^{-\frac{\epsilon}{2}}}g^{(0)}$, where $\alpha_S := \frac{(g^{(\mathrm{DR},\MS)})^2}{4 \pi}$, and $\gamma_E$ is the Euler-Mascheroni constant. It is possible to define an $X$-space scheme strong coupling $\alpha_S^{(\mathrm{DR},X)}$ (for example by fixing the two-point correlation function for the gluon correlator), but this work focuses on the matching of composite operators between $X$-space schemes and $\MS$, and hence the $\alpha_S$ that is used is always renormalized in $\MS$ at scale $\mu$. The $\MS$ renormalization factor 
\begin{equation}
Z_{(\overline{q} \Gamma Q)}^{(\mathrm{DR},\MS)}(\mu)= 1 - \frac{\alpha_S(\mu)}{\pi \epsilon} - \frac{\alpha_S(\mu)}{2\pi} \log (4 \pi e^{-\gamma_E}), 
\end{equation}
can be read off from \cref{eq:BBDRX} as the $\frac{1}{\epsilon}$ piece and the corresponding $\MS$ logarithms. Comparing this to \Cref{eq:BBDRX}, note that the $\MS$-counterterm contains factors of $+\frac{\alpha_S \gamma_E}{\pi}$, whereas the $X$-space counterterm contains corresponding factors of $-\frac{\alpha_S\gamma_E}{\pi}$. The difference in sign arises from the $d$-dimensional Fourier transform of the logarithmic structure \cite{Chetyrkin:2010dx}. This causes the conversion factor $\mathcal{O}^{(\mathrm{DR},\MS)} = C^{(\mathrm{DR},\MS;\mathrm{DR},X)}(t,\mu) \mathcal{O}^{(\mathrm{DR},X)}(t)$ between $X$-space and $\MS$ to contain $\gamma_E$ factors: 
\begin{equation}
C^{(\mathrm{DR},\MS;\mathrm{DR},X)}_{(\overline{q} \Gamma Q)}(t,\mu) = \frac{Z_{(\overline{q} \Gamma Q)}^{(\mathrm{DR},\MS)}(\mu)}{Z^{(\mathrm{DR},X)}_{(\overline{q} \Gamma Q)}(t,\mu)} = 1 + \frac{\alpha_S(\mu)}{\pi} \left( \frac{4}{3} + \frac{2 \pi^2}{9} + \frac{1}{2}\log \left(e^{2 \gamma_E}\mu^2 t^2 \right) \right).
\end{equation}
In numerical studies utilizing $X$-space schemes, $\mu$ should be varied to provide an indication of the size of the error caused by truncating the perturbative series. A natural candidate for a central value of $\mu$ is given by the condition $\mu^2 = e^{-2 \gamma_E} t^{-2}$ which would cancel the logarithm that appears in the matching factor. 

\begin{table}[t]
\centering
\bgroup
\def\arraystretch{1.35}
\begin{tabular}{|l|l|l|}
\hline
$\mathcal{O}$  & $C_{1,\mathcal{O}}$ & $C_{2,\mathcal{O}}$  \\ \hline
$\Lambda_1$ & $\frac{17}{6}$  & $2$ \\ \hline
$\Lambda_2$ & $\frac{9}{4}$ &  $1$ \\ \hline
$\Sigma_{1,\alpha}, \Sigma_{1,\alpha,i}^*$ & $\frac{19}{12}$  & $1$ \\ \hline
$\Sigma_{2,\alpha}, \Sigma_{2,\alpha,i}^*$ & $\frac{11}{6}$ &  $\frac{2}{3}$ \\ \hline
\end{tabular} \hspace{0.5cm}
\begin{tabular}{|l|l|l|}
\hline
$\mathcal{O}$ & $C_{1,\mathcal{O}}$ & $C_{2,\mathcal{O}}$  \\ \hline
$\Lambda_1^-$ & $\frac{3}{2}$ &  $2$ \\ \hline
$\Lambda_2^-, \Lambda_1^{*-}$ & $\frac{9}{4}$ &  $1$ \\ \hline
$\Sigma_{2,\alpha}^-$ & $\frac{19}{12}$  & $1$ \\ \hline
$\Sigma_{1,\alpha}^- , \Sigma_{1,\alpha,i}^{*-}$ & $\frac{11}{6}$  & $\frac{2}{3}$ \\ \hline
\end{tabular}
\egroup
\caption{$O(\alpha_S)$ conversion factors from $X$-space renormalized operators to $\MS$-renormalized operators, for the different heavy-light-light baryonic operators. 
Refer to \Cref{eq:conv} for definitions of $C_{1,\mathcal{O}},C_{2,\mathcal{O}}$. The two-point functions of the positive-parity heavy-light-light currents have previously been computed in the context of QCD sum rules \cite{Grozin:1992td}, but did not appear in position space explicitly.}\label{tab:tab}
\end{table}

For the heavy-light-light baryons, the $O(\alpha_S)$ conversion factors can be parametrized similarly as
\begin{equation}\label{eq:conv}
C^{(\mathrm{DR},\MS; \mathrm{DR},X)}_{\mathcal{O}}(t,\mu) = \frac{Z_{\mathcal{O}}^{(\mathrm{DR},\MS)}(\mu)}{Z^{(\mathrm{DR},X)}_{\mathcal{O}}(t,\mu)} = 1 + \frac{\alpha_S}{\pi} \left( C_{1,\mathcal{O}} +  \frac{2\pi^2}{9} + \frac{C_{2,\mathcal{O}}}{2} \log ( e^{2 \gamma_E} \mu^2 t^2) \right),
\end{equation}
where the coefficients $C_{1,\mathcal{O}},C_{2,\mathcal{O}}$ are given in \Cref{tab:tab} for $\mathcal{O}$ taken from \Cref{eq:lam,eq:sig}.\footnote{Note that $Z^{(\mathrm{DR},\mathrm{MS})}_\mathcal{O}$ (which contains only the $\frac{1}{\epsilon}$ piece) can be read off as $Z^{(\mathrm{DR},\mathrm{MS})}_\mathcal{O} = \frac{-\alpha_S}{\pi \epsilon} C_{2,\mathcal{O}}$ as the $\gamma_E$ factors are directly proportional to the $\frac{1}{\epsilon}$ pole.} 
The calculations of the conversion factors proceed similarly to those for the heavy-light mesonic operators, by evaluating the baryonic analogues to the diagrams appearing in Fig.~\ref{fig:BB}. 

\subsection{Ratios of decay constants}
The renormalization conditions presented in this section are not directly applicable to two-point correlation functions computed with the static heavy-quark action in Lattice-HQET as $\mathcal{O}^{(\mathrm{DR},X)}(t) \neq \mathcal{O}^{(\mathrm{lat},X)}(t)$. The reason for this disagreement is that the lattice regulator introduces a power-divergent mixing between the static kinetic operator $\overline{Q} D_0 Q$ and a radiatively generated mass-term $m_\text{stat} \overline{Q} Q$ where $m_\text{stat} \sim O(\alpha_S)/a$. The relationship between matrix elements of the operators renormalized in the two different regulators is 
\begin{equation}
\langle \cdots | \mathcal{O}^{(\mathrm{lat},X)}(t) | \cdots \rangle = e^{m_\text{stat} t} \langle \cdots | \mathcal{O}^{(\mathrm{DR},X)}(t) | \cdots \rangle, 
\end{equation}
where $| \cdots \rangle$ represents an arbitrary state. Without an additional renormalization condition that can be used to extract $m_\text{stat}$, it is not possible to match matrix elements of $\mathcal{O}^{(\mathrm{lat},X)}$ to matrix elements in continuum renormalization schemes. 

Since the self-energy power divergence affects all the two-point correlation functions discussed in this section in the same fashion, it is, however, possible to take ratios to cancel this self-energy divergence. 
As an example application, consider the following QCD matrix elements:
\begin{align}
\langle 0 | (\overline{q}_f \gamma_\mu \gamma_5 b)^{(\mathrm{DR},\MS)}  | \overline{B}_f(p) \rangle &= i p_\mu f_{B_f}^{\MS}, \label{eq:decay1}\\
\langle 0 |  \left(\epsilon^{abc} [q^{aT} \tau_A C  \gamma_5 q^b] \tfrac{1 + \gamma_0}{2} b^c_\alpha  \right)^{(\mathrm{DR},\MS)} | \Lambda_{b}(p,s) \rangle &= i m_{\Lambda_{b}} N_\alpha(p,s) f_{\Lambda_{b,1}}^{\MS}, \label{eq:decay2}
\end{align}
where $b$ is a (relativistic) bottom quark field, $p_\mu$ is a Euclidean four-momentum, $s$ is the spin of the $\Lambda_b$-baryon, and $N_\alpha$ is the baryon Dirac spinor satisfying $\overline{N}(p,s) N(p,s') = 2 m_{\Lambda_b} \delta_{s s'}$. Here, $f_{B_f}$ is the mesonic decay-constant, and $f_{\Lambda_{b,1}}$ is the normalization constant for one of the distribution amplitudes of the $\Lambda_b$ baryon \cite{Ball:2008fw}. The states in \Cref{eq:decay1,eq:decay2} have standard relativistic normalization
\begin{align}
\langle \overline{B}_f(p') | \overline{B}_f(p) \rangle &= 2 E_{\overline{B}_f(p)} (2 \pi)^3 \delta^3(\vec{p} - \vec{p}'), \\
\langle \Lambda_b(p',s') | \Lambda_b(p,s) \rangle &= 2 \delta_{ss'} E_{\Lambda_b(p,s)} (2 \pi)^3 \delta^3(\vec{p} - \vec{p}').
\end{align}
Note that, given the conventions of \cref{eq:decay1,eq:decay2}, the mass dimension of $f_{B_f}$ is $1$, while the mass dimension of $f_{\Lambda_{b,1}}$ is $2$ (using the normalization for distribution amplitudes from Ref~\cite{QCDSF:2008qtn}). The decay constants $f_{B_f}, f_{\Lambda_{b,1}}$ as defined above do not have well-defined limits as $m_b \to \infty$, as the relativistic normalization of states does not behave well in this limit. The $m_b \to \infty$ limit can be studied by switching to a non-relativistic normalization for the HQET states:
\begin{align}
{}_\mathrm{NR}\langle \overline{B}_f(v,k') | \overline{B}_f(v,k) \rangle_\mathrm{NR} &= (2 \pi)^3 \delta^3(\vec{k} - \vec{k}'), \\
{}_\mathrm{NR}\langle \Lambda_b(v,k',s') | \Lambda_b(v,k,s) \rangle_\mathrm{NR} &= \delta_{ss'} (2 \pi)^3 \delta^3(\vec{k} - \vec{k}'),
\end{align}
where $p_\mu = m_X v_\mu + k_\mu$. The relativistically normalised and non-relativistically normalised states additionally differ by $O(\frac{1}{m_b})$ corrections, which are irrelevant in a static limit analysis. Furthermore, matching the QCD operators to HQET operators with $\MS$ matching factors $D(\mu)$, and dropping the $O(\frac{1}{m_b})$ contributions gives \cite{UKQCD:1993pvm}
\begin{align}
H_{f}^{(\mathrm{DR},\MS)} 
&= (\overline{q}_f \gamma_5 Q)^{(\mathrm{DR},\MS)} 
= D_{H_f}(\mu) (\overline{q}_f \gamma_0 \gamma_5 b)^{(\mathrm{DR},\MS)},\\
\Lambda_{1,\alpha}^{(\mathrm{DR},\MS)} 
&= \left([q^{aT} \tau_A C  \gamma_5 q^b] Q^c_\alpha \epsilon_{abc} \right)^{(\mathrm{DR},\MS)} 
= D_{\Lambda_1}(\mu) \left([q^{aT} \tau_A C  \gamma_5 q^b] \tfrac{1 + \gamma_0}{2} b^c_\alpha \epsilon_{abc} \right)^{(\mathrm{DR},\MS)}.
\end{align}
Combining the matching of the operators and states from QCD to HQET with \Cref{eq:decay1,eq:decay2} gives the relationships
\begin{align}
i m_{B_f} f_{B_f}^{\mathrm{stat},\MS} &= D_{H_f}(\mu)^{-1} \sqrt{2 m_{B_f}} \langle 0 | H_{f}^{(\mathrm{DR},\MS)} | \overline{H}_f(v,k) \rangle_\mathrm{NR}, \label{eq:hqet1}\\ 
i m_{\Lambda_b} N_\alpha(p,s) f_{\Lambda_{b,1}}^{\mathrm{stat},\MS} &= D_{\Lambda_1}(\mu)^{-1} \sqrt{2 m_{\Lambda_{b,1}}} \langle 0 | \Lambda_{1,\alpha}^{(\mathrm{DR},\MS)} | \Lambda_b(v,k,s) \rangle_\mathrm{NR}, \label{eq:hqet2}
\end{align}
where the superscript `stat' has been prepended to the label of the decay constants to emphasize that $O(\frac{1}{m_b})$ corrections have been dropped in the derivation. The matrix elements on the right-hand sides of \cref{eq:hqet1,eq:hqet2} are defined completely in the static HQET limit, and hence have no heavy-quark mass dependence. Therefore, up to logarithmic corrections due to the matching factors, the combinations $f_{B_f}^{\mathrm{stat},\MS} \sqrt{m_{B_f}}$ and $f_{\Lambda_{b,1}}^{\mathrm{stat},\MS} m_{\Lambda_b}$ are constant in the $m_b \to \infty$ limit. 

To extract the ratio of the static decay constants using the $X$-space scheme as an intermediate nonperturbative renormalization scheme, the bare two-point correlation functions are first calculated in Lattice HQET:
\begin{align}
T_{H_f}(t) &:= \langle H_f^{(\mathrm{lat},0)}(t,\vec{0}) H_f^{(\mathrm{lat},0)\dagger} (0,\vec{0}) \rangle = |Z_{H_f}|^2 e^{-E_{H_f} t} + \text{excited states}, \\
T_{\Lambda_{1}}(t) &:= \sum_\alpha \left\langle  \Lambda_{1,\alpha}^{(\mathrm{lat},0)}(t,\vec{0}) \Lambda_{1,\alpha}^{(\mathrm{lat},0)\dagger} (0,\vec{0})  \right\rangle = |Z_{\Lambda_1}|^2 e^{-E_{\Lambda_{Q,1}} t} + \text{excited states},
\end{align}
and fitted at large Euclidean time-separations, $t$, to extract $Z_{H_f}$ and $Z_{\Lambda_1}$. Note that the energies $E_{H_f},E_{\Lambda_1}$ are the binding energies of the respective hadrons shifted by the static quark mass $m_\text{stat}$. By renormalizing the operators in the $X$-space scheme at reference scale $t_0^{-1}$ and then matching to the $\MS$ scheme, an expression for the ratio of the decay constants in the $\MS$ scheme can be derived as
\begin{align}
\frac{f_{B_f}^\mathrm{stat,\MS}}{f_{\Lambda_{b,1}}^{\mathrm{stat,\MS}}}(\mu) 
=
\left( \frac{T_{H_f}(t_0)}{T_{H_f}^{\mathrm{NI}}(t_0)} \frac{T_{\Lambda_{1}}^{\mathrm{NI}}(t_0)}{T_{\Lambda_{1}}(t_0)} \right)^{\frac{1}{2}}
&\times \left( \frac{C^{\mathrm{DR},\MS; \mathrm{DR},X}_{\overline{Q} \Gamma q} (t_0,\mu)}{C^{\mathrm{DR},\MS; \mathrm{DR},X}_{\Lambda_{1}} (t_0,\mu)}  \right) \times \left( \frac{D_{\Lambda_{1}}(t_0)}{D_{H_f}(t_0)} \right) 
\times \frac{Z_{H_f} \sqrt{2} m_{\Lambda_{b}}}{Z_{\Lambda_{1}} \sqrt{m_{B_f}}}
\end{align}
where the first factor renormalizes the bare operators in the $X$-space scheme, the second factor converts to the $\MS$ scheme in the dimensionally-regulated continuum, and the third factor matches the renormalized HQET operators to QCD operators in the $\MS$ scheme. All the renormalization and matching in this expression is performed at the scale $t_0^{-1}$ for simplicity, but hybrid schemes where the renormalized operators are run before performing matching are also possible. For instance, running the operators in $\MS$ from the scale $t_0^{-1}$ to the scale $m_b$ before matching the HQET operators to QCD is usually desirable, as this cancels the large logarithms appearing if the $t_0^{-1}$ scale is very different from $m_b$.

\section{Four-quark Operators}

\subsection{Flavor non-singlet $\Delta Q = 0$ four-quark operators}
\label{sec:db0}

By performing an Operator Product Expansion (OPE) for the product of two weak currents and matching to HQET (a procedure known in the literature as the Heavy Quark Expansion \cite{Lenz:2014jha}), the lifetimes of hadrons containing a heavy quark $Q$ ($Q = c,b$) can be expressed as a sum over matrix elements of operators of increasing dimension. 
Of the various $O(1/m_Q^3)$ corrections, the ``spectator effects"  
arising for four-quark operators where a light spectator quark $q_f$ in the hadron participates in the decay along with the heavy-quark are phase-space enhanced in the OPE by a factor of $16 \pi^2$ \cite{Neubert:1996we}. In HQET, the operators of interest are conventionally written in the basis
\begin{align}\label{eq:4q}
O^f_1 &:=  (\overline{Q} \gamma_\mu P_L q_f)(\overline{q}_f \gamma_\mu P_L Q),  
&O^f_2 &:=  (\overline{Q}  P_L q_f)(\overline{q}_f  P_R Q) \nonumber,\\
O^f_3 &:=  (\overline{Q} \gamma_\mu P_L T^A q_f)(\overline{q}_f \gamma_\mu P_L T^A Q),  
&O^f_4 &:=  (\overline{Q}  P_L T^A q_f)(\overline{q}_f  P_R T^A Q),
\end{align}
where $f \in \{u,d\}$, $P_L = \frac{1 - \gamma_5}{2}, P_R = \frac{1 + \gamma_5}{2}$ are the left/right projectors, and the $T^A$ are color matrices satisfying $\mathrm{Tr}(T^A T^B) = \frac{1}{2} \delta^{AB}$. Note that the flavor-singlet combinations of the operators such as $O^u + O^d$ will mix with the lower-dimensional $\overline{Q} Q$ operator in a power-divergent way. When acting on $B$-hadrons, $\overline{Q}Q$ is the identity operator, leading to an $a^{-3}$ additive mixing to the operators in lattice-HQET computations. This section focuses on the renormalization of the isospin-nonsinglet contributions such as $O^u - O^d$ where the mixing with the $\overline{Q} Q$ operators cancel, and the four-quark operators are protected from power-divergent mixing. The $f$-label on the four-quark operators is omitted in what follows, as it should be understood that all operators refer to the isospin-nonsinglet versions. 

In dimensional regularization, four-quark operators such as those listed in \Cref{eq:4q} mix with evanescent operators, which are operators that formally vanish in $d = 4$ due to their Dirac structure. Different choices of basis for the evanescent operators lead to finite shifts in the $\MS$ renormalized matrix elements \cite{Herrlich:1994kh,Ciuchini:1997bw}. Working in the HV scheme, the basis of evanescent operators appearing at $O(\alpha_S)$ chosen here is written as

\begin{figure}[t]
\includegraphics{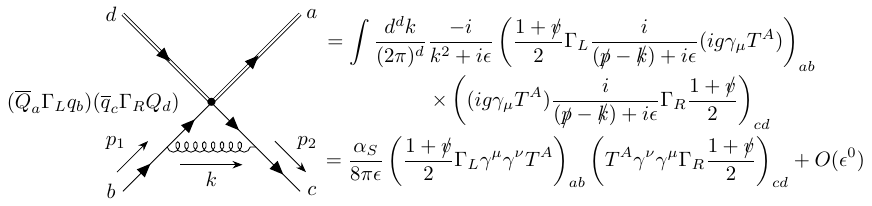}
\caption{The one-loop diagram in Minkowski space for the $\Delta Q = 0$ operators that generates the evanescent structures shown in \Cref{eq:ev}. The indices $a,b,c,d$ are combined Dirac-color indices. }\label{fig:ev}
\end{figure}

\begin{align}\label{eq:ev}
E_1 &:= (\overline{Q} \gamma_\mu P_L \gamma_\alpha \gamma_\beta q) (\overline{q} \gamma_\beta \gamma_\alpha \gamma_\mu P_L Q) - 
4 O_1,  \nonumber \\
E_2 &:= (\overline{Q} P_L \gamma_\alpha \gamma_\beta q) (\overline{q} \gamma_\beta \gamma_\alpha P_R Q) - 4 O_2,  \nonumber \\
E_3 &:= (\overline{Q} \gamma_\mu P_L \gamma_\alpha \gamma_\beta T^A q) (\overline{q} \gamma_\beta \gamma_\alpha \gamma_\mu P_L T^A Q) - 4 O_3,  \nonumber \\
E_4 &:= (\overline{Q} P_L \gamma_\alpha \gamma_\beta T^A q) (\overline{q} \gamma_\beta \gamma_\alpha P_R T^A Q) - 4 O_4.  
\end{align}
The Dirac structures present in these evanescent operators occur in the one-loop diagram with a gluon attached to the two light quarks, as shown in Fig.~\ref{fig:ev}. In order for the proposed $X$-space scheme to be regulator-independent, evanescent contributions to operators must be subtracted for all regulators, and the renormalization conditions must be formulated for the subtracted operators \cite{Lehner:2011fz,Ciuchini:1995cd}. In general, all operators in $\{\mathcal{O}_1,\mathcal{O}_2,\mathcal{O}_3,\mathcal{O}_4\}$ will mix with each other, but to $O(\alpha_S)$ in $\MS$ the operators mix in sub-bases, with $\{\mathcal{O}_1,\mathcal{O}_3,E_1,E_3\}$ having the same mixing pattern as $\{\mathcal{O}_2,\mathcal{O}_4,E_2,E_4\}$. The MS-renormalization conditions (not containing the logarithmic factors included in $\MS$) for $i \in \{1,2\}$ are given by

\begin{equation}\label{eq:evmix}
\begin{pmatrix} O^{(0)}_{i} \\ O^{(0)}_{i+2} \end{pmatrix} =  
{\renewcommand{\arraystretch}{1.5}
\begin{pmatrix} 
1 + \fdfrac{2 \alpha_S}{\pi \epsilon} &\ &
-\fdfrac{3 \alpha_S}{2 \pi \epsilon} &\ &
0 &\ &
\fdfrac{\alpha_S}{8 \pi \epsilon} \\
-\fdfrac{\alpha_S}{3\pi \epsilon} &\ &
1 + \fdfrac{\alpha_S}{4 \pi \epsilon} & \ &
\fdfrac{\alpha_S }{36 \pi \epsilon} & \  &
\fdfrac{7 \alpha_S}{48 \pi \epsilon} 
\end{pmatrix}}   \begin{pmatrix} O^{(\mathrm{MS})}_{i} \\ O^{(\mathrm{MS})}_{i+2} \\ E^{(\mathrm{MS})}_{i} \\ E^{(\mathrm{MS})}_{i+2} \end{pmatrix}.
\end{equation}
The first generation of bare evanescent operators $\{E_1,\cdots,E_4\}$ themselves mix at $O(\alpha_S)$ with a second generation of bare evanescent operators containing even more complicated Dirac structures (such as $(\overline{Q} P_L \gamma_{\alpha_1} \gamma_{\alpha_2} \gamma_{\alpha_3} \gamma_{\alpha_4} q)(\overline{q} \gamma_{\alpha_4} \gamma_{\alpha_3}\gamma_{\alpha_2} \gamma_{\alpha_1} P_R Q)$) \cite{Herrlich:1994kh}. Such higher-order evanescent operators are omitted in \cref{eq:evmix} as the matching conditions presented later between $\MS$ and $X$-space schemes are not sensitive to them at $O(\alpha_S)$. 

Subtracting the $\frac{1}{\epsilon}$ evanescent contributions to the physical operators gives evanescent-subtracted operators $\tilde{\mathcal{O}}_i$ that can be used in regulator-independent schemes. By reading off the coefficients from \cref{eq:evmix} they are defined to be, for $i \in \{1,2\}$,
\begin{equation}
\tilde{{O}}_{i}^{(0)} = {O}_{i}^{(0)} - \frac{\alpha_S}{8 \pi \epsilon} E_{i+2}^{(0)},\quad  \tilde{{O}}_{i+2}^{(0)} = {O}_{i+2}^{(0)} - \frac{\alpha_S}{36 \pi \epsilon} E_{i}^{(0)} - \frac{7 \alpha_S}{48 \pi \epsilon} E_{i+2}^{(0)}.
\end{equation}

As the four-quark operators being considered are $\Delta Q = 0$, and the static quark can only travel in the timelike direction, an $X$-space scheme utilising two-point correlation functions of $\tilde{\mathcal{O}}_i$ (similar to that proposed in \cref{sec:source} for the bilinear and trilinear operators) is not possible to define. The reason is that the corresponding two-point correlation functions are zero (due to the $\theta(-t_E)$ portion of the static heavy quark propagator shown in \Cref{eq:hqprop}). A possible way to rectify this is to compactify the time direction (for instance, in a thermal calculation), but this would likely be significantly more complicated due to the Matsubara sums required in the computation \cite{Bernard:1974bq}. 

Instead, three-point correlation functions combining $\tilde{\mathcal{O}}_i$ with different choices of source and sink operators can be used to define an $X$-space renormalization scheme. Such an approach was also considered, for example, in the $X$-space renormalization of the QCD stress-energy tensor \cite{Costa:2021iyv}. Taking ratios of three-point correlation functions to appropriate two-point correlation functions cancels the renormalization factors of the source and sink operators, provided they are multiplicatively renormalizable. In HQET, this has the added benefit of cancelling the static-quark self-energy divergence.  
Writing the renormalized $\Delta Q = 0$ operators as $\tilde{\mathcal{O}}^{(R,X)}_i(t) = Z^{(R,X)}_{ij}(t) \tilde{\mathcal{O}}^{(0)}_j$,  
the renormalization condition is defined here by a choice of four source-operator/sink-operator combinations (labelled as $J_n,K_n$ respectively, $n \in \{1,2,3,4\}$) and is given by
\begin{equation}
\frac{\langle J_n^{\dagger}(-t,\vec{0}) \tilde{\mathcal{O}}_i^{(R,X)}(0,\vec{0}) K^{}_n(t,\vec{0})\rangle }{\sqrt{\big|\langle J_n^{\dagger}(-t,\vec{0}) J_n^{}(t,\vec{0}) \rangle \langle K_n^{\dagger}(-t,\vec{0}) K_n^{}(t,\vec{0}) \rangle \big|}} 
= 
\frac{\langle J_n^{\dagger}(-t,\vec{0}) \tilde{\mathcal{O}}^{(0)}_i(0,\vec{0}) K_n(t,\vec{0})\rangle }{\sqrt{\big| \langle J_n^{\dagger}(-t,\vec{0}) J_n(t,\vec{0})  \rangle \langle K_n^{\dagger}(-t,\vec{0}) K_n(t,\vec{0})  \rangle\big|}} \Bigg|_\mathrm{NI},
\end{equation}
(for all $n$; no sum over $n$ implied) at a fixed $t$. In the same way as for the two-point $X$-space condition presented in \Cref{sec:source}, additional indices on the source and sink operators should be summed over on both sides of the condition, and the open spinor index is traced over if the source and sink are baryonic. Since the source and sink operators are multiplicatively renormalizable, this causes the $Z$-factors of the source and sink to cancel in these ratios, so they are not labelled as bare or renormalized. Furthermore, the static-quark self-energy cancels in these ratios of correlation functions with the same physical length of the Wilson line, allowing for nonperturbative renormalization of the operators without determination of $m_\mathrm{stat}$.
Defining
\begin{equation}
T_{i,n}(t) := \frac{\langle J_n^{\dagger}(-t,\vec{0}) \tilde{\mathcal{O}}_i^{(0)}(0,\vec{0}) K^{}_n(t,\vec{0})\rangle }{\sqrt{\big| \langle  J_n^{\dagger}(-t,\vec{0}) J_n^{}(t,\vec{0}) 
 \rangle \langle  K_n^{\dagger}(-t,\vec{0}) K_n^{}(t,\vec{0}) 
 \rangle\big|}} = T_{i,n}^{(0)}(t) + \alpha_S T_{i,n}^{(1)}(t) + O(\alpha_S^2),
\end{equation}
where $T^{(0)}_{i,n}(t)$ is the noninteracting value, the $X$-space renormalization conditions can then be solved as
\begin{equation}\label{eq:ZX}
Z_{ij}^{(X)} = \sum_n T_{i,n}^{(0)}(t) T^{-1}_{n,j}(t) = \mathbbm{1}_{i,j} - \alpha_S  \sum_{n} T^{(1)}_{i,n}(t) (T^{(0)}(t))^{-1}_{n,j} + O(\alpha_S^2),
\end{equation}
which is well-defined as long as the four source/sink pairs, $n \in \{1,2,3,4\}$, are chosen so that the noninteracting matrix $T^{(0)}_{i,n}(t)$ is invertible in $d = 4$.  

\begin{table}[t]
\begin{equation*}
\hspace*{-0.85cm}
\def\arraystretch{1.35}
\begin{array}{|c|c|c||c|c|c|c|c|c|c|c|}
 \hline n & J_{n} & K_n & T_{1,n}^{(0,0)} & T_{2,n}^{(0,0)} & T_{3,n}^{(0,0)} & T_{4,n}^{(0,0)}
 & T_{1,n}^{(0,1)} & T_{2,n}^{(0,1)} & T_{3,n}^{(0,1)} & T_{4,n}^{(0,1)}
 \\\hline
 &&&&&&&&&& \\[-19pt]
 \hline 
 1 & H_f^- & H_f^- &-6&-6&0&0&3&3&0&0 \\ \hline
 2 & H_{f,i}^{*-} & H_{f,i}^{*-} &-6&0&0&0&3&0&0&0 \\ \hline
 3 & \Lambda_1 & \Sigma_{2,0} & -2\sqrt{3}&-\sqrt{3}&\frac{4}{\sqrt{3}}&\frac{2}{\sqrt{3}}&\frac{4}{\sqrt{3}}&\frac{1}{\sqrt{3}}&\frac{-8}{3\sqrt{3}}&\frac{-2}{3\sqrt{3}} \\ \hline
 3^* & \Lambda_2 & \Sigma_{1,0} &2\sqrt{3}&\sqrt{3}&\frac{-4}{\sqrt{3}}&\frac{-2}{\sqrt{3}}&-2\sqrt{3}&\frac{-2}{\sqrt{3}}&\frac{4}{\sqrt{3}}&\frac{4}{3 \sqrt{3}} \\ \hline
 4 & \Sigma_{1,\alpha} & \Sigma_{1,\alpha} &
 6&1&-4&\frac{-2}{3}&\frac{-11}{3}&\frac{-3}{2}&\frac{22}{9}&1 \\ \hline
 4^* & \Sigma_{2,\alpha} & \Sigma_{2,\alpha} &
 -6&-1&4&\frac{2}{3}&\frac{-5}{3}&\frac{-1}{2}&\frac{10}{9}&\frac{1}{3} \\ \hline
 5 & \Sigma_{1,\alpha,i}^* & \Sigma_{1,\alpha,i}^* &0&-2&0&\frac{4}{3}&\frac{2}{3}&0&\frac{-4}{9}&0 \\ \hline
 5^* & \Sigma_{2,\alpha,i}^* & \Sigma_{2,\alpha,i}^* &0&-2&0&\frac{4}{3}&\frac{2}{3}&0&\frac{-4}{9}&0 \\ \hline
\end{array}
\end{equation*}
\caption{Decomposition for the noninteracting ratio of correlation functions according to \Cref{eq:NI_parm}, for varying source-sink pairs $(J_n,K_n)$. 
The source/sink pairs $n = 3$ and $n = 3^*$ give the same noninteracting matrix elements ($T^{(0,0)}_{i,n}$) in $d = 4$ up to a sign, and hence may not both be chosen as part of the set of four source/sink operators used in the renormalization condition due to the requirement that $T^{(0)}_{i,n}$ is invertible. The same is true for $n = 4,4^*$ and $n = 5,5^*$. }\label{tab:NI_rat}
\end{table}

The natural candidates for the source and sink operators are the mesonic and baryonic currents discussed in \cref{sec:source}. 
The requirement that $T^{(0)}_{i,n}(t)$ is invertible means that it is not possible to use four mesonic source/sink pairs, as, for any mesonic source/sink pair $(J_M,K_M)$, the matrix element $\langle J_M^\dagger (-t,\vec{0}) \tilde{\mathcal{O}}^{(0)}_{i} K_M(t,\vec{0}) \rangle|_{\mathrm{NI}}$ vanishes for $i \in \{3,4\}$ due to the color trace. 
Chiral symmetry, heavy quark symmetry, and spin representations cause many source/sink choices to give vanishing matrix elements with all the operators, further restricting the number of distinct choices. 
For the remaining nonzero ratios of correlation functions, the noninteracting ratios are parametrized as
\begin{equation}\label{eq:NI_parm}
T_{i,n}^{(0)}(t) = T_{i,n}^{(0,0)} \frac{1}{\pi^2 t^{3-\epsilon}} \left( \frac{\pi}{4} e^{\gamma_E} \right)^{\frac{\epsilon}{2}} + T_{i,n}^{(0,1)} \frac{\epsilon}{\pi^2 t^{3-\epsilon}}  + O(\epsilon^2).
\end{equation}
In dimensional regularization, after removing factors of $\delta^{d-1}$ corresponding to the $\delta$-function in position space from the static quark propagators, $T_{i,n}(t)$ has dimension $3-\epsilon$, accounted for by the factor of $t^{-(3-\epsilon)}$ in \Cref{eq:NI_parm}. The specific source/sink pairs that are studied in this section are the negative-parity heavy-light mesonic operators and the positive-parity heavy-light-light baryonic operators discussed in \Cref{sec:source}. The values of the decomposition for the noninteracting ratio for these source/sink operators $(J,K)$ are tabulated in \Cref{tab:NI_rat}. Every choice of four linearly independent source/sink operators from this list (there are $28$ different choices in total) defines a different $X$-space scheme. In a similar way to the parametrization of the noninteracting contribution to the ratio of correlation functions in \Cref{eq:NI_parm}, the $O(\alpha_S)$-contribution to the ratios are parametrized by
\begin{equation}\label{eq:OA_decomp}
T_{i,n}^{(1)} =  T_{i,n}^{(1,0)} \frac{1}{\epsilon \pi^3 t^{3-2\epsilon}} \left(\frac{\pi}{4} e^{\gamma_E} \mu \right)^\epsilon + T_{i,n}^{(1,1)} \frac{1 }{\pi^3 t^{3-2\epsilon}} + T_{i,n}^{(1,2)} \frac{1}{\pi t^{3-2\epsilon}} + O (\epsilon).
\end{equation}
For the various source/sink pairs, the $O(\alpha_S)$ matrix elements for the ratio have been calculated and are tabulated in \Cref{tab:OAS}. 

\begin{table}[t]
\begin{equation*}
\def\arraystretch{1.35}
\begin{array}{|c||c|c|c|c|c|c|c|c|c|c|c|c|}
\hline n 
 & T_{1,n}^{(1,0)} & T_{2,n}^{(1,0)} & T_{3,n}^{(1,0)} & T_{4,n}^{(1,0)}
 & T_{1,n}^{(1,1)} & T_{2,n}^{(1,1)} & T_{3,n}^{(1,1)} & T_{4,n}^{(1,1)}
 & T_{1,n}^{(1,2)} & T_{2,n}^{(1,2)} & T_{3,n}^{(1,2)} & T_{4,n}^{(1,2)}
 \\ \hline  &&&&&&&&&&&& \\[-19pt] \hline 
 1  &-12&-12&2&2&-10&-10&\frac{4}{3}&\frac{5}{6}&-\frac{8}{3}&-\frac{8}{3}&-\frac{4}{9}&-\frac{4}{9}  \\ \hline 
 2  &-12&0&2&0&-10&0&\frac{2}{3}&\frac{1}{6}&-\frac{8}{3}&0&-\frac{4}{9}&0 \\ \hline
 {3}  & -6 \sqrt{3}&-3 \sqrt{3}&\sqrt{3}&\frac{\sqrt{3}}{2}&-\frac{8}{\sqrt{3}}&-\frac{7}{\sqrt{3}}&\frac{7}{3 \sqrt{3}}&\frac{5}{3 \sqrt{3}}&\frac{4}{3 \sqrt{3}}&\frac{2}{3 \sqrt{3}}&\frac{10}{9 \sqrt{3}}&\frac{5}{9 \sqrt{3}} \\ \hline
 {3^*}  &6 \sqrt{3}&3 \sqrt{3}&-\sqrt{3}&-\frac{\sqrt{3}}{2}&\sqrt{3}&\frac{3\sqrt{3}}{2}&-\frac{2}{\sqrt{3}}&-\frac{\sqrt{3}}{2}&-\frac{4}{3\sqrt{3}}&-\frac{2}{3 \sqrt{3}}&-\frac{10}{9 \sqrt{3}}&-\frac{5}{9 \sqrt{3}} \\ \hline
 {4}  &18&3&-3&-\frac{1}{2}&\frac{22}{3}&-1&-\frac{43}{18}&\frac{5}{12}&-\frac{4}{3}&-\frac{2}{9}&-\frac{10}{9}&-\frac{5}{27} \\ \hline
 4^*  & -18&-3&3&\frac{1}{2}&-\frac{73}{3}&-\frac{31}{6}&\frac{103}{18}&\frac{25}{36}&\frac{4}{3}&\frac{2}{9}&\frac{10}{9}&\frac{5}{27}  \\ \hline
 5  &0&-6&0&1&\frac{4}{3}&-6&-\frac{8}{9}&\frac{3}{2}&0&\frac{4}{9}&0&\frac{10}{27} \\ \hline
 5^*  &0&-6&0&1&\frac{4}{3}&-\frac{19}{3}&-\frac{8}{9}&\frac{31}{18}&0&\frac{4}{9}&0&\frac{10}{27} \\ \hline
\end{array}
\end{equation*}
\caption{Decomposition for the $O(\alpha_S)$ contribution to the ratios of correlation functions defined in \Cref{eq:OA_decomp}. The source/sink pair index $n$ is the same as used in \Cref{tab:NI_rat}. }\label{tab:OAS}
\end{table}

Choosing the specific source/sink pairs $n \in \{1,2,3,4\}$ from the list of source/sink pairs presented in \Cref{tab:NI_rat,tab:OAS}, the $X$-space renormalization matrix can be calculated using \Cref{eq:ZX} as
\begin{align}\label{eq:Z_lifetime}
&Z^{(X)}_{ij,n \in \{1,2,3,4\}} = \mathbbm{1} - \nonumber \\
&\qquad \alpha_S(\mu) \left(\frac{\pi}{4} e^{\gamma_E} t^2 \mu^2\right)^{\frac{\epsilon}{2}} \cdot 
\begingroup
\renewcommand*{\arraystretch}{1.35}
\begin{pmatrix}
\frac{2}{\pi \epsilon} + \frac{24 + 4 \pi^2}{9 \pi} & 0 & \frac{-3}{2 \pi \epsilon} + \frac{-9 + 24 \pi^2}{24 \pi} & \frac{-5}{4 \pi} \\
0 & \frac{2}{\pi \epsilon} + \frac{24 + 4 \pi^2}{9 \pi} & \frac{-1}{16 \pi} & \frac{-3}{2 \pi \epsilon} + \frac{-63 + 72 \pi^2}{72 \pi} \\
\frac{-1}{3 \pi \epsilon} + \frac{-15 + 4 \pi^2}{54 \pi} & \frac{-1}{9 \pi} & \frac{1}{4 \pi \epsilon} + \frac{90 + 56 \pi^2}{144 \pi} & \frac{-1}{12 \pi} \\
\frac{-1}{36 \pi} & \frac{-1}{3 \pi \epsilon} + \frac{-15 + 4 \pi^2}{54 \pi} & \frac{-3}{16 \pi} & \frac{1}{4 \pi \epsilon} + \frac{414 + 168 \pi^2}{432 \pi}
\end{pmatrix}
\endgroup,
\end{align}
where $\mathbbm{1}$ is the $4 \times 4$ identity matrix. As a check, it can be verified that the $\frac{1}{\epsilon}$ divergent pieces match up with the $\mathrm{MS}$ counterterms presented in \Cref{eq:evmix}. The corresponding conversion factor from the $X$-space scheme to $\MS$ is given by
\begin{align}\label{eq:matchdb0}
&C^{(\MS;X)}_{ij,n \in \{1,2,3,4\}} := \sum_k Z^{(\MS)}_{ik} (Z^{(X)})^{-1}_{kj,n \in \{1,2,3,4\}} = \mathbbm{1} + \nonumber\\
&\qquad \alpha_S(\mu)\left(
\renewcommand*{\arraystretch}{1.35}
\begin{array}{cccc}
 \frac{\log (\beta )}{\pi }+\frac{4 \pi }{9}+\frac{8}{3 \pi } & 0 & -\frac{3 \log (\beta )}{4 \pi }+\pi -\frac{3}{8 \pi } & -\frac{5}{4 \pi } \\
 0 & \frac{\log (\beta )}{\pi }+\frac{4 \pi }{9}+\frac{8}{3 \pi } & -\frac{1}{16 \pi } & -\frac{3 \log (\beta )}{4 \pi }+\pi -\frac{7}{8 \pi } \\
 -\frac{\log (\beta )}{6 \pi }+\frac{2 \pi }{27}-\frac{5}{18 \pi } & -\frac{1}{9 \pi } & \frac{\log (\beta )}{8 \pi }+\frac{7 \pi }{18}+\frac{5}{8 \pi } & -\frac{1}{12 \pi } \\
 -\frac{1}{36 \pi } & -\frac{\log (\beta )}{6 \pi }+\frac{2 \pi }{27}-\frac{5}{18 \pi } & -\frac{3}{16 \pi } & \frac{\log (\beta )}{8 \pi }+\frac{7 \pi }{18}+\frac{23}{24 \pi }
   \\
\end{array}
\right),
\end{align}
where $\beta := e^{2 \gamma_E} \frac{\mu^2 t^2}{16}$. A natural choice for $\mu$ is $\mu^2 = 16e^{-2\gamma_E}/t^2$ which would cancel the factors of $\log(\beta)$ appearing in the matching coefficient. In principle, when converting matrix elements computed with lattice HQET to $\MS$-renormalized matrix elements, varying over the different choices of source/sink pairs, as well as varying over the scale $t$ that the $X$-space scheme is defined at before perturbatively running to a common scale, will give an indication on the error due to $O(\alpha_S^2)$ terms that have been neglected in this study.

\subsection{$\Delta Q = 2$ four-quark operators}\label{sec:db2}

In the Standard Model, neutral $B$-mesons ($B^0, B^0_s$) mix with their own antiparticles. The lowest-order diagram contributing to this in the Standard Model is a box diagram with the exchange of two weak bosons, which after integrating out physics at and above the weak scale leads to $\Delta B = 2$ four-quark operators. When matching these operators to HQET, the Lagrangian is expanded to include a static antiquark $Q_-$ that travels in the opposite direction to the static quark $Q_+$ (compare to the $Q_+$-propagator \Cref{eq:hqprop}):
\begin{equation}\label{eq:antihqprop}
\langle Q_-^{(0)}(0) \overline{Q}_-^{(0)}(x_E) \rangle_F = \delta_{\vec{x}_E,\vec{0}} \ \theta(t_E)  W^{(0)}(0,x_E) \frac{1 - \gamma_0}{2},
\end{equation}
where $\frac{1 \pm \gamma_0}{2} Q_\pm = Q_\pm$. A basis of the relevant $\Delta Q = 2$ operators in HQET is
\begin{equation}\label{eq:db0op}
\begin{aligned}
&O_1 := (\overline{Q}_+  P_L q) (\overline{Q}_-  P_L q), 
&O_2 := (\overline{Q}_+ P_L T^A q) (\overline{Q}_- P_L T^A q), \\
&O_3 := (\overline{Q}_+ P_L q)(\overline{Q}_- P_R q),
&O_4 := (\overline{Q}_+ P_L T^A q) (\overline{Q}_- P_R T^A q).
\end{aligned}
\end{equation}
The operators $O_1,O_2$ contribute to neutral $B$-meson mixing in the Standard Model, and $O_3,O_4$ are contributions that could arise from possible new physics \cite{Gabbiani:1996hi}. The full basis of $\Delta Q = 2$ operators also contains $\{O_1,O_2\}$ with left-handed projectors swapped with right-handed projectors ($P_L \leftrightarrow P_R$), but these are related by parity and so renormalize with the same factors as $\{O_1,O_2\}$. There are also additional $\Delta Q = 2$ operators when $Q$ is relativistic, but these are related to the operators in \Cref{eq:db0op} in the static quark limit \cite{Becirevic:2001xt}:
\begin{equation}
(\overline{Q}_+ \gamma_\mu P_L q) (\overline{Q}_- \gamma_\mu P_L q) = \frac{8}{3} O_1 + 4 O_2 \quad \quad \mathrm{in}\quad d = 4.
\end{equation}
The evanescent operators are defined as
\begin{align}
E_1 &:= (\overline{Q}_+ P_L \gamma_\alpha \gamma_\beta q)(\overline{Q}_- P_L \gamma_\alpha \gamma_\beta q) - \frac{32}{3} O_1 - 16 O_2, \nonumber \\  
E_2 &:= (\overline{Q}_+ P_L T^A \gamma_\alpha \gamma_\beta) (\overline{Q}_- P_L T^A \gamma_\alpha \gamma_\beta q) - \frac{32}{9} O_1 - \frac{16}{3} O_2 \nonumber \\
E_3 &:= (\overline{Q}_+ P_L \gamma_\alpha \gamma_\beta q)(\overline{Q}_- P_R \gamma_\alpha \gamma_\beta q) - 4 O_3, \nonumber \\ 
E_4 &:= (\overline{Q}_+ P_L T^A \gamma_\alpha \gamma_\beta) (\overline{Q}_- P_R T^A \gamma_\alpha \gamma_\beta q) - 4 O_4,
\end{align}
where the $\gamma$-matrix structure is governed by the $O(\alpha_S)$ diagram with a single gluon attached to the two light quarks. The mixing pattern is given by the following MS renormalization:
\begin{equation}\label{eq:db2_eq1}
\begin{pmatrix} O_{1}^{(0)} \\ O_{2}^{(0)} \end{pmatrix} = 
{\renewcommand{\arraystretch}{2}\begin{pmatrix} 1 + \fdfrac{14 \alpha_S}{9 \pi \epsilon} & \fdfrac{4 \alpha_S}{3 \pi \epsilon} & 0 & -\fdfrac{\alpha_S}{8 \pi \epsilon} \\ \fdfrac{8 \alpha_S}{27 \pi \epsilon} & 1 + \fdfrac{10 \alpha_S}{9 \pi \epsilon} & \fdfrac{-\alpha_S}{36 \pi \epsilon} & \fdfrac{\alpha_S}{24 \pi \epsilon} \end{pmatrix}} 
\begin{pmatrix} O_1^{\mathrm{MS}} \\ O_{2}^{\mathrm{MS}} \\ E_1^{\mathrm{MS}} \\ E_2^{\mathrm{MS}} \end{pmatrix}, 
\end{equation}
\begin{equation}\label{eq:db2_eq2}
\begin{pmatrix} O_{3}^{(0)} \\ O_{4}^{(0)} \end{pmatrix} = 
{\renewcommand{\arraystretch}{2}\begin{pmatrix} 1 + \fdfrac{2 \alpha_S}{ \pi \epsilon} & \fdfrac{3 \alpha_S}{2 \pi \epsilon} & 0 & -\fdfrac{\alpha_S}{8 \pi \epsilon} \\ \fdfrac{\alpha_S}{3 \pi \epsilon} & 1 + \fdfrac{3 \alpha_S}{2 \pi \epsilon} & \fdfrac{-\alpha_S}{36 \pi \epsilon} & \fdfrac{\alpha_S}{24 \pi \epsilon} \end{pmatrix}} 
\begin{pmatrix} O_3^{\mathrm{MS}} \\ O_{4}^{\mathrm{MS}} \\ E_3^{\mathrm{MS}} \\ E_4^{\mathrm{MS}} \end{pmatrix}.
\end{equation}
Correspondingly, the evanescent-subtracted operators are defined for $i \in \{1,3\}$ as
\begin{equation}
\tilde{O}_i^{(0)} = O_i^{(0)} + \frac{\alpha_S}{8 \pi \epsilon} E_{i+1}^{(0)}, \quad \tilde{O}_{i+1}^{(0)} = \tilde{O}_{i+1}^{(0)} + \frac{\alpha_S}{36 \pi \epsilon} E_i^{(0)} - \frac{\alpha_S}{24 \pi \epsilon} E_{i+1}^{(0)} 
\end{equation}

Baryonic heavy-light-light currents cannot be used as source/sink pairs for the $\Delta B = 2$ four-quark operators, as the corresponding three-point functions all vanish. Fortunately, enough constraints can be derived with the mesonic heavy-light currents as source/sink pairs to constitute a valid $X$-space scheme. Although chiral symmetry is formally broken by the HV $\gamma_5$ scheme, the massless nature of the light quarks causes the operators to mix in the $2 \times 2$ subblocks presented in \Cref{eq:db2_eq1,eq:db2_eq2}. Thus, only two source/sink pairs are needed in the $X$-space scheme. Using the choice $(\overline{Q} \gamma_5 q, \overline{q} \gamma_5 Q)$ and $(\overline{Q} \gamma_i \gamma_5 q, \overline{q} \gamma_i \gamma_5 Q)$, the $O(\alpha_S)$ matching matrices are found to be
\begin{equation}\label{eq:matchdb20}
\hspace{1.2cm}C^{(\MS,X)}_{\{O_1,O_2\}} = \mathbbm{1} + \frac{\alpha_S}{\pi} \left(
{\renewcommand{\arraystretch}{1.5}\begin{array}{ccc}
 \fdfrac{7 \log (\beta)}{9}+\fdfrac{4 \pi ^2}{9}+\fdfrac{23}{9} &\ &
 \fdfrac{2 \log(\beta)}{3}-\fdfrac{\pi ^2}{3}+\fdfrac{4}{3} \\
 \fdfrac{4 \log (\beta)}{27}-\fdfrac{2 \pi ^2}{27}+\fdfrac{8}{27} &\ &
 \fdfrac{5 \log (\beta) }{9 }+\fdfrac{5 \pi ^2}{9}+\fdfrac{19}{9} \\
\end{array}}
\right),
\end{equation}

\begin{equation}\label{eq:matchdb21}
C^{(\MS,X)}_{\{O_3,O_4\}} = \mathbbm{1} + \frac{\alpha_S}{\pi} \left(
{\renewcommand{\arraystretch}{1.5}\begin{array}{ccc}
 \log(\beta) +\fdfrac{4 \pi ^2}{9}+\fdfrac{25}{9} &\ &
 \fdfrac{3\log(\beta)}{4}-\fdfrac{\pi ^2}{3}+\fdfrac{7}{6} \\
 \fdfrac{\log(\beta)}{6}-\fdfrac{2 \pi ^2}{27}+\fdfrac{7}{27} &\ &
 \fdfrac{3\log(\beta)}{4}+\fdfrac{5 \pi ^2}{9}+\fdfrac{43}{18} \\
\end{array}}
\right),
\end{equation}
where $\beta := e^{2 \gamma_E} \frac{\mu^2 t^2}{16}$. 

\section{Conclusion}\label{sec:conc}

In this work, a set of $X$-space renormalization schemes for isospin-nonsinglet $\Delta Q = 0$ and $\Delta Q = 2$ four-quark HQET operators have been proposed, and the $O(\alpha_S)$ matching coefficients between these schemes and $\MS$ in the dimensionally regulated continuum have been calculated. This allows for a gauge-invariant, nonperturbative renormalization matrix elements calculated in lattice HQET, without the need to extract the power-divergent self-energy contribution $m_\mathrm{stat}$. 
Precise computations of these matrix elements with lattice HQET will reduce theory uncertainties on lifetimes of heavy hadrons, and help constrain physics beyond the Standard Model. 
Note that, when implementing the $X$-space scheme for these four-quark operators in lattice HQET, it is convenient to use Ginsparg-Wilson discretizations of the light quarks (e.g., with the domain-wall fermion action) to avoid additional mixing between the four-quark operators of interest and operators in other chiral representations.   

Next-to-next to leading order calculations of the matching coefficients presented in \Cref{eq:matchdb0,eq:matchdb20,eq:matchdb21} at $O(\alpha_S^2)$ are possible, but the computation is complicated by the fact that, unlike $p$-type integrals that only have one external scale, the perturbative calculations shown in \Cref{subsec:int} have two external scales $x_\mathrm{src},x_\mathrm{snk}$ corresponding to the source and sink locations of the three-point renormalization scheme proposed. Corrections due to finite light-quark masses are more easily calculable (though they are likely smaller than the $O(\alpha_S^2)$ corrections on typical lattice-QCD ensembles), and require computation of the three-loop integrals in \Cref{subsec:int} either analytically in the light-quark mass, or by expanding in powers of the light quark mass. $O(\frac{1}{m_Q})$ corrections are in principle also calculable, but require considering mixing of the four-quark operators with dimension-7 operators that contain an additional covariant derivative, as well as considering the $O(\frac{1}{m_b})$ corrections to the static HQET lagrangian. 

The calculations and techniques used in this work can be readily applied to $X$-space schemes for other classes of operators. For instance, using the auxiliary-field formalism, nonlocal operators such as $\overline{q}(x) W(x,y) q(y)$, where $W(x,y)$ is a Wilson line, are transformed into products of local operators $\overline{q}Q_{y - x}(x) \overline{Q}_{y-x} q(y)$, which can be renormalized by the techniques presented in \Cref{sec:source}. The calculations of the three-loop diagrams involving gluons attached to the light-quark propagators presented in \Cref{app:charge} can also be applied to renormalize massless four-quark operators such as the $\Delta S = 1$ four-quark operators relevant for kaon decays. 

\acknowledgments

The authors gratefully acknowledge useful discussions with Artur Avkhadiev and Aditya Pathak. S.M.\ is supported by the U.S. Department of Energy, Office of Science, Office of High Energy Physics under Award Number DE-SC0009913.
W.D. and J.L. are supported in part by the
U.S. Department of Energy, Office of Science under grant Contract Number DE-SC0011090 and by the SciDAC5 award DE-SC0023116 and are additionally supported by the National Science Foundation under Cooperative Agreement PHY-2019786 (The NSF AI Institute for Artificial Intelligence and Fundamental Interactions, http://iaifi.org/). 

\appendix

\section{Integrals and Conventions}\label{sec:app}

\subsection{Dimensional regularization and charge conjugation}
\label{app:charge}
The calculations in this work use dimensionally regulated integrals $ \int \frac{d^4 k}{(2 \pi)^4} \to \mu^{4-d} \int \frac{d^d k}{(2 \pi)^d}$ where $d = 4 - \epsilon$. 
$\gamma_5$ is treated as in the 't Hooft-Veltman scheme \cite{tHooft:1972tcz} ($\gamma_5 := i \gamma_0 \gamma_1 \gamma_2 \gamma_3$ in Minkowski signature). 
Charge-conjugation matrices $C$ are used in the construction of the baryonic operators such as $\epsilon^{abc} [q^{aT} C \Gamma q^b] Q^c$ (where $\Gamma$ is a Dirac matrix), and it is natural to assume that the defining relation $C \gamma_\mu C^{-1} = -\gamma_\mu^T$ holds in dimensional regularization.
However, to our knowledge, an explicit charge-conjugation matrix satisfying the defining relations for an explicit basis of infinite-dimensional $\gamma$-matrices has not been constructed previously in the literature for dimensional regularization (a construction was presented for dimensional reduction in Ref~\cite{Stockinger:2005gx}). 
The explicit construction shows that enforcing $C \gamma_\mu C^{-1} = -\gamma_\mu^T$ does not lead to inconsistencies in Dirac traces, unlike how naively enforcing the anticommuting relation $\{\gamma_5,\gamma_\mu\} = 0$ leads to inconsistencies in certain Dirac traces. 

In what follows, a construction of the $d$-dimensional gamma matrices as well as an explicit charge-conjugation matrix $C$ satisfying $C \gamma_\mu C^{-1} = \gamma_\mu^T$ are presented.
The Minkowski-signature gamma matrices are defined inductively following the algorithm of Ref~\cite{Collins:1984xc} (up to a trivial reordering that enforces $(\overline{\gamma}^\mu)^T = (-1)^\mu \overline{\gamma}^\mu$):

\begin{itemize}
\item[1)] Set
\begin{equation}
\gamma^0_{(1)} := \begin{bmatrix} 1 & 0 \\ 0 & -1 \end{bmatrix}, \quad \gamma^1_{(1)} := \begin{bmatrix} 0 & 1 \\ -1 & 0 \end{bmatrix}.
\end{equation}
\item[2)] For $\omega \in \mathbb{Z}_{\geq 1}$, define
\begin{equation}
\hat{\gamma}_{(\omega)} := i^{\omega - 1} \gamma^0_{(\omega)} \cdots \gamma^{2 \omega - 1}_{(\omega)},
\end{equation}
\begin{equation}
\gamma^\mu_{(\omega + 1)} := \begin{bmatrix} \gamma^\mu_{(\omega)} & 0 \\ 0 & \gamma^\mu_{(\omega)} \end{bmatrix} \quad \text{for} \quad 0 \leq \mu < 2 \omega,
\end{equation}
\begin{equation} 
  \gamma^{2 \omega}_{(\omega + 1)} := \begin{bmatrix} 0 & i \hat{\gamma}_{(\omega)} \\ i \hat{\gamma}_{(\omega)} & 0 \end{bmatrix} , \quad \gamma^{2 \omega + 1}_{(\omega + 1)} := \begin{bmatrix} 0 & \hat{\gamma}_{\omega} \\ - \hat{\gamma}_{(\omega)} & 0 \end{bmatrix}. 
\end{equation}
\item[3)] The infinite-dimensional $\gamma$-matrices are defined by block-diagonal copies of the finite-dimensional construction, so that for $\mu \in \mathbb{Z}_{\geq 0}$, choosing any $\omega \geq \lfloor \frac{\mu}{2} \rfloor + 1$,
\begin{equation}
  \overline{\gamma}^\mu := \begin{bmatrix}
    \gamma^\mu_{(\omega)} & & \\
    & \gamma^\mu_{(\omega)} & \\
    & & \ddots 
  \end{bmatrix}. 
\end{equation}
\end{itemize}
The $\overline{\gamma}$ matrices defined here satisfy
\begin{equation}\label{eq:G}
\{ \overline{\gamma}_\mu , \overline{\gamma}_\nu \} = 2 g_{\mu \nu}, \quad \overline{\gamma}_\mu^\dagger = \overline{\gamma}_0 \overline{\gamma}_\mu \overline{\gamma}_0, 
\end{equation}
where the metric is written in the mostly-negative convention, $g_{\mu \nu} = \text{diag}(+1,-1,-1,\cdots)$. 
With this construction, no finite product of $\overline{\gamma}$ matrices will satisfy the charge-conjugation matrix condition $C \overline{\gamma}_\mu C^{-1} = -\overline{\gamma}_\mu^T$. Modifying the basis of $\gamma$-matrices by eliminating $\gamma_4,\gamma_6,\gamma_8\dots$ from the basis, such that the new basis $\gamma_\mu$ is given by the relabelling
\begin{equation}
\{{\gamma}_0,{\gamma}_1,{\gamma}_2,{\gamma}_3,{\gamma}_4,{\gamma}_5,{\gamma}_6,\cdots\} = \{\overline{\gamma}_0,\overline{\gamma}_1,\overline{\gamma}_2,\overline{\gamma}_3,\overline{\gamma}_5,\overline{\gamma}_7,\overline{\gamma}_9,\cdots\},
\end{equation}
a charge-conjugation matrix can be defined. These $\gamma$ matrices still satisfy \cref{eq:G}, but now we can define $C = i \gamma_0 \gamma_2$, which satisfies
\begin{equation}
C \gamma_\mu^T C^{-1} = -\gamma_\mu^T, \quad C \gamma_5 C^{-1} = \gamma_5, \quad C^{-1} = C^T = C^\dagger = -C.
\end{equation}
The trace is normalized such that $\mathrm{Tr}(\mathbbm{1}) = 4$. Euclidean $\gamma$-matrices are obtained by defining
$\gamma_0^\text{E} := \gamma_0, \gamma_i^\text{E} := -i \gamma_i$ such that $\{\gamma_\mu^E, \gamma_\nu^E\} = 2 \delta_{\mu \nu}$. The `E' labels are dropped from all Euclidean $\gamma$-matrices in the main text as all calculations are presented in Euclidean space.

\subsection{Integrals}\label{subsec:int}

\begin{figure}
\centering
\includegraphics{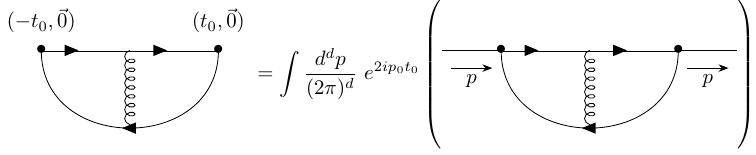}
\caption{Example showing how the Fourier-transform relates a position-space diagram appearing in the $\langle \overline{Q} \Gamma q (t_0,\vec{0}) (\overline{q} \overline{\Gamma} Q) (-t_0,\vec{0})\rangle $ two-point correlation function to the corresponding momentum-space diagram. }\label{fig:fourier}
\end{figure}

Calculations were performed in Mathematica, with the aid of the Tracer package \cite{Jamin:1991dp} for Dirac traces in the `t Hooft-Veltman scheme, and HypExp \cite{Huber:2005yg} for expansions of hypergeometric functions. Because three-quark baryonic sources were used in defining the $X$-space schemes (which only exist in $N_c = 3$, and introduce an $\epsilon^{abc}$ tensor), color traces are explicitly evaluated without attempting to write in terms of $N_c$. Color matrices are normalised so that $\mathrm{Tr}(T^A T^B) = \frac{1}{2} \delta^{AB}$. 

The two-point correlation functions required for renormalizing the heavy-light and heavy-light-light operators from \Cref{sec:source} can be computed first in momentum space and then by taking a Fourier transform, as shown in Fig.~\ref{fig:fourier}. As such, these can be evaluated by using `$p$-type' integrals \cite{Grozin:2003ak}. In the chiral limit for the light quark, there is only a single dimensionful scale $p$ in the momentum-space integral corresponding to the momenta running through the diagram as seen in Fig.~\ref{fig:fourier}. Hence, integration-by-parts relations based on the identity $\int \frac{d^d k}{(2 \pi)^d} \frac{\partial}{\partial k_\mu} f(k) = 0$ are relatively easy to derive. 

For three-point position-space functions, at $O(\alpha_S)$, the diagrams can all be split up into a number of component pieces. For example, one diagram topology that appears in the $O(\alpha_S)$ contribution to the three-point mesonic function $\langle 
 \overline{Q} \Gamma^\dagger q (-t_0,\vec{0}) \cdot  (\overline{Q} \Gamma_L q)(\overline{q} \Gamma_R Q) (0,\vec{0}) \cdot \overline{q} \Gamma Q (t_0,\vec{0}) \rangle $ can be written as

\begin{center}
\includegraphics{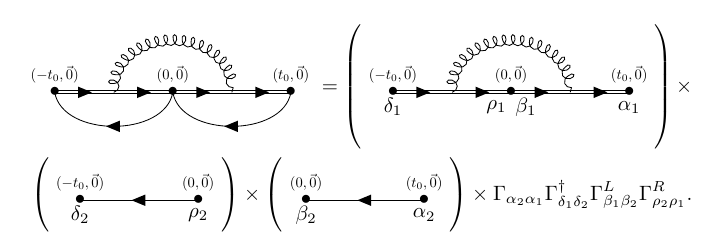}
\end{center}

Here, $\alpha_i, \beta_i, \rho_i, \delta_i$ are Dirac-color indices. As well as the position-space propagators and self-energy diagrams, there are three $O(\alpha_S)$ diagrams to compute, corresponding to a gluon attaching to two heavy-quark propagators, a gluon attaching on one end to a heavy-quark propagator and on the other end to a light-quark propagator, and a gluon attaching to two light-quark propagators. In the case of the $\Delta Q = 0$ four-quark operators, the diagram with a gluon attaching to two heavy-quark propagators can be directly calculated in Minkowski space as follows:

\includegraphics{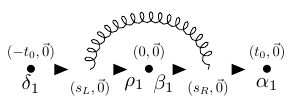}
\begin{align}\label{eq:hh}
  &= \left[\PP (i g v_\mu T_A) \right]_{\alpha_1 \beta_1} \left[\PP (ig v_\mu T_A) \right]_{\rho_1 \delta_1}  \cdot \int_{-t_0}^0 ds_L \int_0^{t_0} ds_R \int \frac{d^d k}{(2 \pi)^d} \frac{-i e^{i s_L k - i s_R k}}{k^2}  \nonumber \\ 
    & = [\slashed{v} T_A ]_{\alpha_1 \beta_1} [\slashed{v} T_A ]_{\rho_1 \delta_1} \left( -\frac{\alpha_S}{\pi \epsilon} - \frac{\alpha_S}{2 \pi} \left(2 + \log \left(- \frac{1}{4} e^{\gamma_E} \pi \mu^2 t_0^2\right) \right) \right), 
\end{align}
where the static nature of the heavy quark has been utilized to integrate the vertex insertions at $s_L$ and $s_R$ along the line connecting the three operators. For the two diagrams involviong gluons attaching to light-quarks, additional master integrals are required. Relabelled from the basis from Appendix A of \cite{Costa:2021iyv}, and in Minkowski space, there is a Tripod diagram $T$ and a Wedge diagram $W$:

\begin{center}
\includegraphics{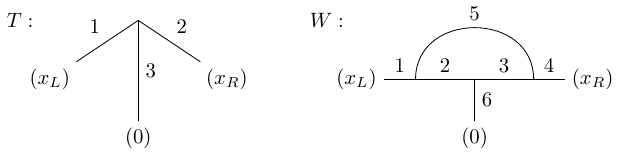}
\end{center}

\begin{equation}
T(x_L, x_R; n_1,n_2,n_3) := \int \frac{d^d p_L d^d p_R}{(2 \pi)^{2d}} \frac{e^{i p_L x_L} e^{- i p_R x_R}}{(-p_L^2)^{n_1} (-(p_L - p_R)^2)^{n_2} (-p_R^2)^{n_3}},
\end{equation}
\begin{equation}\label{eq:W}
\begin{split}
&W(x_L,x_R; n_1,n_2,n_3,n_4,n_5,n_6) := \\
&\quad \int \frac{d^d p_L d^d p_R d^d k}{(2 \pi)^{3d}} \frac{e^{i p_L x_L} e^{- i p_R x_R} 
}{(-p_L^2)^{n_1} (-(p_L - k)^2)^{n_2} (-(p_R - k)^2)^{n_3} (-p_R^2)^{n_4} (-k^2)^{n_5} {(-(p_L-p_R)^2)^{n_6}}}.\\
  \end{split}
\end{equation}
The $W$ master integral can be reduced to a few base cases by use of integration-by-parts relations (derived from inserting $\partial_k \cdot k$ in front of the integrand of \cref{eq:W}):
\begin{equation}\label{eq:rec}
W(x_L,x_R; \vec{n}) = \frac{n_2 2^+ (5^- - 1^-) + n_3 3^+ (5^- - 4^-)}{d - n_2 - n_3 - 2n_5} W(x_L,x_R; \vec{n}),
\end{equation}
where $\vec{n} = (n_1,\cdots,n_6)$, and $m^\pm W(x_L,x_R,\vec{n}) = W(x_L,x_R,\vec{n}')$ with $\vec{n}' = \vec{n}$ for all components except the $m$-th component, $n'_m = n_m \pm 1$ (this is the notation used in Ref~\cite{Grozin:2003ak}). \Cref{eq:rec} reduces the $W$ master integral to base cases where either $n_2,n_3$ or $n_5$ equals zero in the argument of $W$. In these cases, the integral reduces to a $p$-type integral and the $T$ master integral, which can be performed explicitly using Schwinger parameters:
\begin{equation}
  \begin{split}
&T(x_L,x_R;n_1,n_2,n_3) = \frac{-\Gamma(\frac{d}{2}-n_1) \Gamma(d - n_1 - n_2 - n_3)}{\Gamma(n_2) \Gamma(n_3) \Gamma(\frac{d}{2}) 4^{n_1 + n_2 + n_3} \pi^d} (-x_R^2)^{-d + n_1 + n_2 + n_3} \\
&\int_0^1 \mathrm{d}x (1-x_1)^{-\frac{d}{2} + n_1 + n_2 - 1} x_1^{-\frac{d}{2} + n_1 + n_3 - 1} {}_2F_1\left(\frac{d}{2} - n_1, d - n_1 - n_2 - n_3, \frac{d}{2}, \frac{- (x_L - x_1 x_R)^2}{x_1 (1 - x_1) x_R^2} \right). 
  \end{split}
\end{equation}
Explicit evaluations of $T$ can be performed at relevant values of $n_1,n_2$ and $n_3$. Diagrams with a gluon attaching to a heavy-quark propagator on one end and to a light-quark propagator on the other end can be calculated in Minkowski space, for instance:

\includegraphics{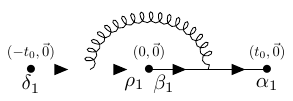}
\begin{align}\label{eq:hl}
&= \int \frac{d^d p_L d^d p_R}{(2\pi)^{2d}} \int_{-t_0}^0 ds_L \left[\frac{i}{\slashed{p}_R} (i g \gamma_\mu T_A) \frac{i}{\slashed{p}_R - \slashed{p}_L} \right]_{\alpha_1 \beta_1} \left[\frac{1 + \slashed{v}}{2}(i g v_\mu T^A) \right]_{\rho_1 \delta_1} \frac{-i e^{i p_L^0 s_L - i p_R^0 t_0}}{p_L^2} \nonumber \\
&= -i g^2 \int_{-t_0}^0 ds_L  \left[ \gamma_\alpha \slashed{v} \gamma_\beta T^A \right]_{\alpha_1 \beta_1} \left[ \frac{1 + \slashed{v}}{2} T^A\right]_{\rho_1 \delta_1} \frac{\partial}{\partial x_R^\alpha} \left(\frac{\partial}{\partial x_R^\beta} + \frac{\partial}{\partial x_L^\beta}\right) T(x_L,x_R; 1,1,1) \bigg|_{\substack{x_L \to (s_L,\vec{0}) \\ x_R \to (t_0, \vec{0})}} 
\end{align}
Finally, the diagram with a gluon attaching to two light lines can be reduced to the $W$ master integral

\begin{align}\label{eq:ll}
  &\adjincludegraphics[valign=c]{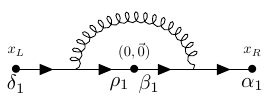} = -i g^2 \mu^{4-d} (\gamma^\alpha \gamma^\mu \gamma^\beta T^a)_{\alpha \beta} (\gamma^\rho \gamma^\mu \gamma^\delta T^a)_{\rho \delta} \nonumber \\
&\qquad \qquad \quad \times \int  \frac{d^d p_L d^d p_R d^d k}{(2 \pi)^{3d}} \frac{e^{i p_L x_L - i p_R x_R} \ p_R^\alpha (p_R - k)^\beta (p_L - k)^\rho p_L^\delta}{(-p_L^2)(-(p_L-k)^2)(-(p_R-k)^2)(-p_R^2)(-k^2)}, 
\end{align}
where the factors of $p_L, p_R$ in the numerator can be handled by differentiating with respect to $x_L, x_R$. Calculating \cref{eq:hh,eq:hl,eq:ll} at the relevant values of $x_L$ and $x_R$ is the main computation involved in calculating the $O(\alpha_S)$ contribution to the ratios of three-point correlation functions to two-point correlation functions presented in \cref{tab:OAS}.

% Bibliography

%% [A] Recommended: using JHEP.bst file
\bibliographystyle{JHEP}
\bibliography{biblio.bib}
% \end{thebibliography}
\end{document}